\documentclass[fleqn,10pt]{wlscirep}
\usepackage[utf8]{inputenc}
\usepackage[T1]{fontenc}

\usepackage{graphicx} 

\usepackage{amsmath,amsthm,amssymb,mathrsfs}
\usepackage{bm}

\usepackage{color}

\title{Bifurcation to complex dynamics in largely modulated voltage-controlled parametric oscillator}

\author[1*]{Tomohiro Taniguchi}
\affil[1]{National Institute of Advanced Industrial Science and Technology (AIST), Research Center for Emerging Computing Technologies, Tsukuba, Ibaraki 305-8568, Japan}

\affil[*]{tomohiro-taniguchi@aist.go.jp}

\begin{abstract}
An experimental demonstration of a parametric oscillation of a magnetization in a ferromagnet was performed recently by applying a microwave voltage, indicating the potential to be applied in a switching method in non-volatile memories. 
In the previous works, the modulation of a perpendicular magnetic anisotropy field produced by the microwave voltage was small compared with an external magnetic field pointing in an in-plane direction. 
A recent trend is, however, opposite, where an efficiency of the voltage controlled magnetic anisotropy (VCMA) effect is increased significantly by material research and thus, the modulated magnetic anisotropy field can be larger than the external magnetic field. 
Here, we solved the Landau-Lifshitz-Gilbert equation numerically and investigated the magnetization dynamics driven under a wide range of the microwave VCMA effect. 
We evaluated bifurcation diagrams, which summarize local maxima of the magnetization dynamics. 
For low modulation amplitudes, the local maximum is a single point because the dynamics is the periodic parametric oscillation. 
The bifurcation diagrams show distributions of the local maxima when the microwave magnetic anisotropy field becomes larger than the external magnetic field. 
The appearance of this broadened distribution indicates complex dynamics such as chaotic and transient-chaotic behaviors, which were confirmed from an analysis of temporal dynamics. 
\end{abstract}

\begin{document}

\flushbottom
\maketitle
%
%

Voltage controlled magnetic anisotropy (VCMA) effect \cite{weisheit07} modulates a perpendicular magnetic anisotropy at a ferromagnetic metal/nonmagnetic insulating layer interface by modulating of electron states near the interface \cite{duan08,nakamura09,tsujikawa09} and/or inducing magnetic moments \cite{miwa17}. 
It enables us to manipulate the direction of the magnetization in a ferromagnet electrically without the Joule heating, and thus, is expected to be a new writing method in magnetoresistive random access memory (MRAM), whose writing scheme currently relies on spin-transfer torque \cite{slonczewski96,berger96}. 
The material researches for highly efficient VCMA effect has been reported, where the perpendicular magnetic anisotropy is largely modulated by a small voltage \cite{maruyama09,shiota09,nozaki10,endo10,wang11,nozaki13,okada14,skowronski15,nozaki17,nozaki18}. 
The efficiency recently achieved reaches to about 300 fJ/(Vm) \cite{nozaki20}, which corresponds to a magnetic anisotropy field on the order of kilo-Oersted for typical VCMA-based MRAM. 
At the same time, analyses on the magnetization dynamics driven by the VCMA effect have been investigated both experimentally and numerically \cite{shiota12,kanai12,amiri13,grezes16,shiota17,lee17,deng17,song18,wu20,shao22}. 
It has been revealed that the switching is unstable when the pulse width of the voltage is short because the dynamics becomes very sensitive to the pulse shape in such condition \cite{lee17}. 
For a long-pulse regime, however, the switching also becomes unstable due to noise \cite{shiota17}. 


To overcome the issue, a parametric oscillation of the magnetization by applying microwave voltage was proposed in Ref. \cite{yamamoto20}. 
For the magnetization switching in the VCMA-based MRAM using a perpendicularly magnetized free layer, an external magnetic field $H_{\rm appl}$ pointing in an in-plane direction is applied and induces the magnetization precession around $H_{\rm appl}$, which eventually relaxes to the direction of $H_{\rm appl}$. 
When the microwave voltage witn an oscillation frequency $f=2 f_{\rm L}$, where $f_{\rm L}=\gamma H_{\rm appl}/(2\pi)$ ($\gamma$ is the gyromagnetic ratio) is the Larmor frequency, is applied, however, a sustainable oscillation of the magnetization is excited. 
This oscillation is classified into the parametric oscillation, and for simplicity, we call the oscillator as voltage-controlled parametric oscillator. 
Since this oscillation is stable, it can be used to manipulate the magnetization direction even in a long-pulse regime, which results in a reliable magnetization switching. 
Note that the previous work \cite{yamamoto20} focuses on a parameter region of $H_{\rm Ka}/H_{\rm appl}\ll 1$, where $H_{\rm Ka}$ is the amplitude of the modulated magnetic anisotropy field generated by the micowave VCMA effect.  
The recent progress of the VCMA efficiency, however, makes the opposite limit, $H_{\rm Ka}/H_{\rm appl}\gg 1$, available because the value of $H_{\rm Ka}$ is growing rapidly, as mentioned above. 
The dynamical behavior of the magnetization in this limit has not been investigated yet. 


In this work, we study the magnetization dynamics driven by the microwave VCMA effect by solving the Landau-Lifshitz-Gilbert (LLG) equation numerically. 
We change the value of $H_{\rm Ka}$ in a wide range including the limit of $H_{\rm Ka}/H_{\rm appl}\gg 1$. 
To clarify the change of the dynamical behavior, bifurcation diagrams are evaluated, which summarize local maxima of the oscillating magnetization. 
When the magnitude of $H_{\rm Ka}$ is small, the parametric oscillation is induced, and the bifurcation diagram becomes single points because the dynamics is periodic. 
When $H_{\rm Ka}$ becomes larger than $H_{\rm appl}$, however, the bifurcation diagrams show broad distributions. 
It is revealed that the appearance of these complex, broadened structures indicated chaotic or transient-chaotic behavior, which are confirmed from evaluations of the Lyapunov exponent and analyses on temporal dynamics. 




\begin{figure}
\centerline{\includegraphics[width=1.0\columnwidth]{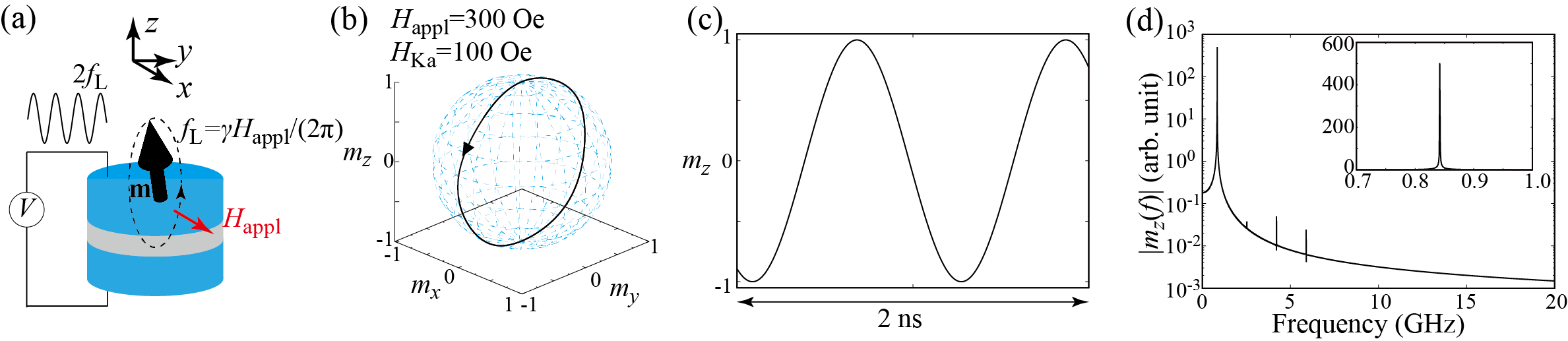}}
\caption{
         (a) Schematic illustration of magnetization oscillation in a ferromagnetic/nonmagnetic trilayer. 
             The unit vector $\mathbf{m}$ pointing in the magnetization direction in free layer shows a sustainable oscillation around an external magnetic field $H_{\rm appl}$ in the in-plane direction when the frequency of a microwave voltage is twice the Larmor frequency $f_{\rm L}$. 
         (b) Dynamical trajectory of the magnetization in a parametric oscillation state.
              Parameters are $H_{\rm appl}=300$ Oe and $H_{\rm Ka}=100$ Oe. 
              The black triangle indicates the direction of the magnetization motion. 
         (c) Time evolution of $m_{z}$ in the parametric oscillation state. 
         (d) Fourier spectrum of $m_{z}$. 
              The inset shows the spectrum around the main peak in a linear scale. 
         \vspace{-3ex}}
\label{fig:fig1}
\end{figure}



\section*{System description}
\label{sec:System description} 

In Fig. \ref{fig:fig1}(a), we show a schematic illustration of a ferromagnetic/nonmagnetic/ferromagnetic trilayer. 
The top and bottom ferromagnets correspond to free and reference layer, respectively. 
We apply a macrospin assumption in the free layer, whose validity in dynamical state driven by the VCMA effect has been confirmed experimentally \cite{yamamoto20}. 
Thus, the unit vector $\mathbf{m}$ pointing in the magnetization direction in the free layer conserves its magnitude, i.e., $|\mathbf{m}|=1$ and $d|\mathbf{m}|/dt=0$. 
An external magnetic field $H_{\rm appl}$ is applied to an in-plane direction. 
We assume that the shape of the free layer is a cylinder, and therefore, the free layer does not have an in-plane magnetic anisotropy. 
For convenience, we use a Cartesian coordinate, where the $x$ axis is parallel to $H_{\rm appl}$ while the $z$ axis is normal to the film plane. 
The magnetic field in the free layer $\mathbf{H}$ , in the absence of an applied voltage, is given by 
\begin{equation}
  \mathbf{H}
  =
  H_{\rm appl}
  \mathbf{e}_{x}
  +
  \left[
    H_{\rm K}
    -
    4\pi M 
    (N_{z}-N_{x})
  \right]
  m_{x}
  \mathbf{e}_{z},
  \label{eq:field_orig}
\end{equation}
where $\mathbf{e}_{k}$ ($k=x,y,z$) is the unit vector in the $k$ direction. 
The perpendicular magnetic anisotropy field includes the contribution from the interfacial magnetic anisotropy field $H_{\rm K}$ \cite{yakata09,ikeda10,kubota12} and the shape magnetic anisotropy field $4\pi M (N_{z}-N_{x})$ with the demagnetization coefficients $N_{k}$ ($N_{x}=N_{y}$ due to the cylindrical symmetry). 
The net perpendicular magnetic anisotropy field, $H_{\rm K}-4\pi M(N_{z}-N_{x})$, determines the retention time of MRAM. 
In the conventional scheme of the writing in the VCMA-based MRAM, the direct voltage modulates this net perpendicular magnetic anisotropy close to zero to excite the magnetization precession around the external magnetic field \cite{shiota16}. 
If the perpendicular component largely remains finite, the magnetization just moves its direction to the direction of the external field with the angle $\sin^{-1}(H_{\rm appl}/H_{\rm K}^{\prime})$ and the precession cannot be excited, where $H_{\rm K}^{\prime}$ is the reduced perpendicular magnetic anisotropy field by the VCMA effect. 
In fact, the numerical simulation in Ref. \cite{shiota16} assumes that the net perpendicular magnetic anisotropy field is completely canceled by the VCMA effect. 
Note that this assumption is important not only for making the simulation simple but also for experiments. 
If the direct (or intrinsic) component of the perpendicular magnetic anisotropy field remains finite during the precession, an instantaneous frequency becomes nonuniform. 
Such a nonuniform frequency will increase the switching error because the pulse width of the voltage for writing the bit is determined as a half of the Larmor precession period for VCMA-based MRAM. 
Therefore, it is preferable to make the direct component of the perpendicular magnetic anisotropy field zero during the switching for the conventional switching scheme. 
In the parametric oscillation state, both direct and microwave voltages are applied to the trilayer, and the direct voltage modulates the total perpendicular magnetic anisotropy so that it becomes close to zero \cite{yamamoto20}, while the microwave voltage provides an oscillating magnetic anisotropy field. 
Accordingly, the magnetic field used in the following calculation becomes, 
\begin{equation}
  \mathbf{H}
  =
  H_{\rm appl}
  \mathbf{e}_{x}
  +
  H_{\rm Ka}
  \sin (2\pi f t)
  m_{z}
  \mathbf{e}_{z},
  \label{eq:field}
\end{equation}
where $H_{\rm Ka}$ and $f$ are the magnitude and frequency of the magnetic anisotropy field due to the microwave voltage. 
The magnetization dynamics driven by this magnetic field is described by the LLG equation, 
\begin{equation}
  \frac{d \mathbf{m}}{dt}
  =
  -\gamma
  \mathbf{m}
  \times
  \mathbf{H}
  +
  \alpha 
  \mathbf{m}
  \times
  \frac{d\mathbf{m}}{dt}, 
  \label{eq:LLG}
\end{equation}
where $\alpha$ is the damping constant. 
Throughout this paper, we use the values of $\gamma=1.764\times 10^{7}$ rad/(Oe s) and $\alpha=0.005$. 
The preparation of the initial state of $\mathbf{m}$ by investigating thermal equilibrium is explained in Methods \cite{taniguchi22,taniguchi23}. 
Recall that the LLG equation conserves the norm of $\mathbf{m}$ as $|\mathbf{m}|=1$. 
Therefore, although $\mathbf{m}$ is a three-dimensional vector, its dynamical degree of freedom is two due to this constraint. 
In fact, if we use a spherical coordinate, for example, the dynamics of $\mathbf{m}$ is described by two variables (zenith and azimuth angles). 
It should be noted that dynamical systems described by differential equations cannot show chaotic behavior when the dynamical degree of freedom is less than or equal to two, according to the Poincar\'e-Bendixson theorem \cite{strogatz01}. 
The presence of the microwave voltage, however, makes the present system non-autonomous and provides a possibility to excite chaotic behavior, as shown below. 


\section*{Results}
\label{sec:Results}


Here, we study the change of the magnetization dynamics for various magnitude of $H_{\rm Ka}$. 


\subsection*{Parametric oscillation}
\label{sec:Parametric oscillation}

Let us first start by confirming the parametric oscillation studied previously \cite{yamamoto20}. 
It was shown in Ref. \cite{yamamoto20} that a sustainable oscillation of the magnetization is excited when the frequency of the microwave voltage, $f$, is twice the Larmor frequency, $f_{\rm L}=\gamma H_{\rm appl}/(2\pi)$. 
Therefore, in the following, we fix the value of $f$ to be $f=2f_{\rm L}$. 
Figure \ref{fig:fig1}(b) shows the dynamical trajectory of the magnetization in a steady state, where $H_{\rm appl}=300$ Oe while $H_{\rm Ka}=100$ Oe, i.e., $H_{\rm Ka}/H_{\rm appl}\ll 1$, as in the case of the previous work \cite{yamamoto20}. 
Since $|\mathbf{m}|=1$ is satisfied in the LLG equation, it is useful to draw the dynamical trajectory on a unit sphere, as shown in this figure. 
Time evolution of $m_{z}$ in a steady state is also shown in Fig. \ref{fig:fig1}(c).
These results indicate an appearance of the sustainable oscillation of the magnetization mentioned above. 
In Fig. \ref{fig:fig1}(d), the Fourier spectrum of $m_{z}$ is shown, where the inset shows it around the main peak in a linear scale. 
Its main peak appears at $0.84$ GHz, which is the same with $f_{\rm L}$ with $H_{\rm appl}=300$ Oe. 
These results are consistent with the previous works \cite{yamamoto20}. 
Since the magnetization switches its direction between $m_{z}\simeq+1$ and $m_{z}\simeq -1$ periodically with the period $1/(2f_{\rm L})$, this parametric oscillation can be used as a switching scheme in VCMA-based MRAM \cite{yamamoto20}. 
Recall that this oscillation is sustained by the microwave modulation of the magnetic anisotropy; if this time-dependent modulation is absent, the magnetization monotonically relaxes to the direction of the external magnetic field. 
Spin-wave propagation through a parametric excitation is another example of the magnetization dynamics caused by microwave voltage, which has been studied previously \cite{verba14,verba16,chen17,rana17}. 


\subsection*{Appearance of complex dynamics and bifurcation diagram}
\label{sec:Appearance of complex dynamics and bifurcation diagram}


\begin{figure}
\centerline{\includegraphics[width=1.0\columnwidth]{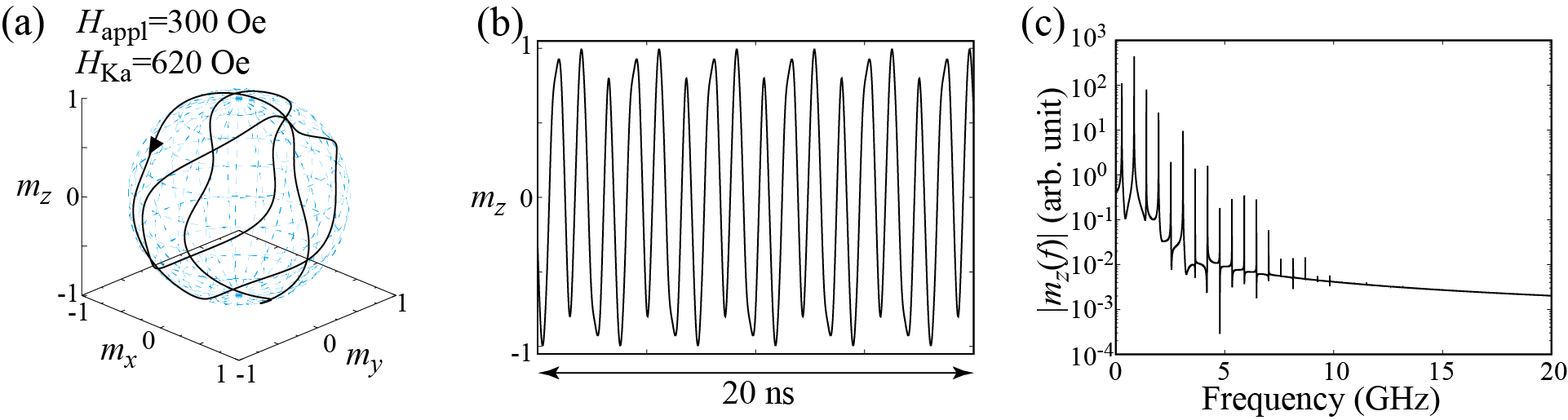}}
\caption{
         (a) Dynamical trajectory of the magnetization in a parametric oscillation state.
              Parameters are $H_{\rm appl}=300$ Oe and $H_{\rm Ka}=620$ Oe. 
              The black triangle indicates the direction of the magnetization motion. 
         (b) Time evolution of $m_{z}$ in the parametric oscillation state. 
         (c) Fourier spectrum of $m_{z}$. 
         \vspace{-3ex}}
\label{fig:fig2}
\end{figure}


When the value of $H_{\rm Ka}$ further increases, the magnetization dynamics becomes complex. 
In Fig. \ref{fig:fig2}(a), we show the dynamical trajectory of the magnetization for $H_{\rm Ka}=620$ Oe, while $H_{\rm appl}=300$ Oe is the same with that used in Fig. \ref{fig:fig1}(b). 
We observe a clear change of the magnetization dynamics by comparing Figs. \ref{fig:fig1}(b) and \ref{fig:fig2}(a). 
The trajectory is not a simple circle in Fig. \ref{fig:fig2}(a). 
The time evolution of $m_{z}$ and its Fourier transformation are shown in Figs. \ref{fig:fig2}(b) and \ref{fig:fig2}(c), respectively. 
Figure \ref{fig:fig2}(b) indicates that the magnetization dynamics is still periodic, while Fig. \ref{fig:fig2}(c) indicates the appearance of multipeak structure. 
These results indicate that the application of the microwave voltage is no longer applicable to the switching method for the VCMA-based MRAM when the modulation of the magnetic anisotropy field becomes larger than the external magnetic field due to the breakdown of the simple parametric oscillation. 


\begin{figure}
\centerline{\includegraphics[width=1.0\columnwidth]{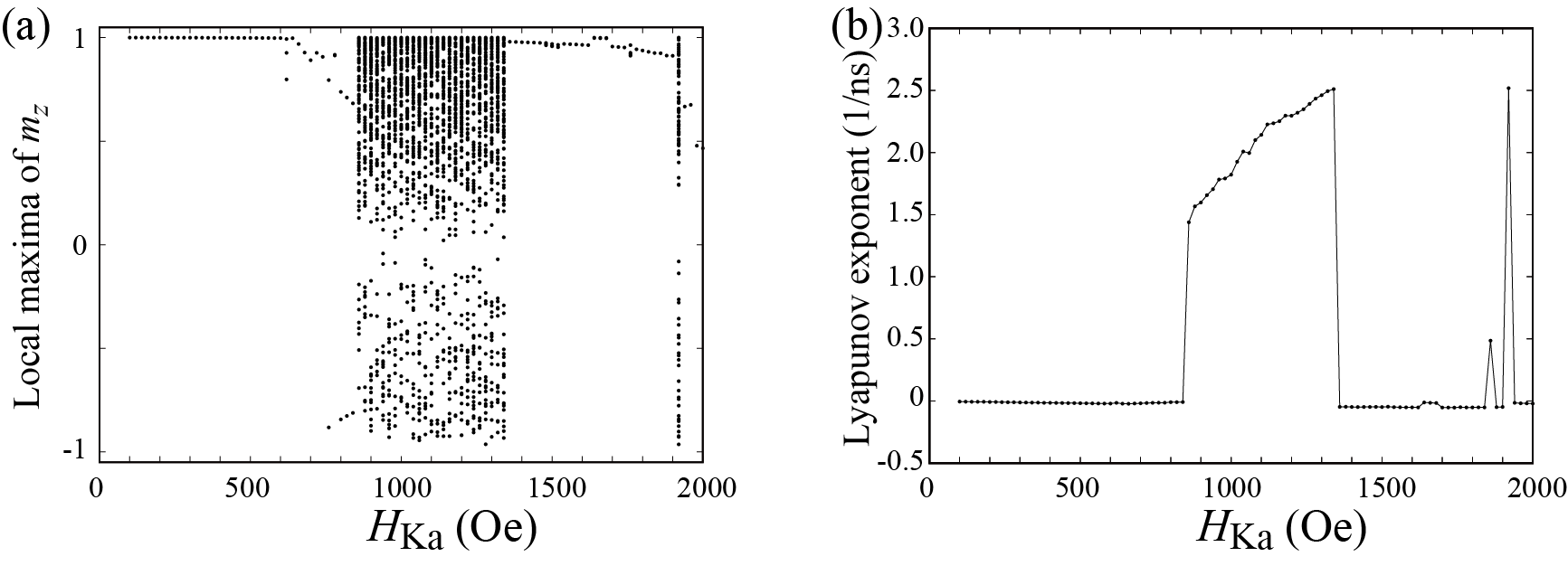}}
\caption{
            (a) A bifurcation diagram summarizing local maxima of $m_{z}(t)$ and (b) Lyapunov exponent for various $H_{\rm Ka}$, where $H_{\rm appl}=300$ Oe. 
         \vspace{-3ex}}
\label{fig:fig3}
\end{figure}


A way to distinguish these dynamics, such as the difference between Figs. \ref{fig:fig1}(b) and \ref{fig:fig2}(a), qualitatively is to draw a bifurcation diagram \cite{strogatz01}. 
In Fig. \ref{fig:fig3}(a), we summarize local maxima of $m_{z}$ for various $H_{\rm Ka}$, where $H_{\rm appl}=300$ Oe. 
Recall that the parametric oscillation is excited when $H_{\rm Ka}$ is small. 
In this case, the dynamics is periodic and $m_{z}$ is similar to a simple trigonometric function, as shown in Fig. \ref{fig:fig1}(c). 
The local maxima of $m_{z}$ for this case, thus, saturate to a single point. 
When the dynamics become complex, the bifurcation diagram shows broadened structure. 
For example, there are three points at $H_{\rm Ka}=620$ Oe in Fig. \ref{fig:fig3}(a), the validity of which is confirmed from Fig. \ref{fig:fig2}(b). 
When the value of $H_{\rm Ka}$ further increases, the bifurcation diagram shows largely broadened structure and finally shows simplified structure again. 
As discussed below, these correspond to chaotic and transient-chaotic behaviors \cite{alligood97,ott02,lai11}. 
Note that the appearance of complex but still periodic structure might weakly depend on the initial state (see Methods) while the region of the broadened structure might depend on the measurement time, as will be mentioned below. 
It is difficult to analytically estimate the value of $H_{\rm Ka}$ at which the bifurcation from a simple parametric oscillation to a complex oscillation occurs; see Methods.  
However, the numerical simulations for various parameters imply that the complex oscillation appears when $H_{\rm Ka}$ becomes larger than $H_{\rm appl}$, as discussed below. 
We also evaluated Lyapunov exponent \cite{strogatz01} by Shimada-Nagashima method \cite{shimada79}, as shown in Fig. \ref{fig:fig3}(b). 
The Lyapunov exponent is an inverse of a time scale characterizing an expansion of a difference between two solutions of the LLG equation with an infinitesimally different initial conditions; see Methods explaining the evaluation method of the Lyapunov exponent. 
A negative Lyapunov exponent means that the magnetization saturates to a fixed point. 
The Lyapunov exponent is zero when the magnetization dynamics are periodic, while it becomes positive when the dynamics are chaotic. 
We notice that the Lyapunov exponent in the present system is zero or positive, depending on the value of $H_{\rm Ka}$. 
Thus, the Lyapunov exponent can be used as an indicator to distinguish between periodic oscillation and chaotic dynamics. 
For example, we can conclude that the dynamics in Fig. \ref{fig:fig2}(a) is periodic not only from the temporal dynamics shown in Fig. \ref{fig:fig2}(b) but also from the fact that the Lyapunov exponent for $H_{\rm Ka}=620$ Oe is zero. 



\begin{figure}
\centerline{\includegraphics[width=1.0\columnwidth]{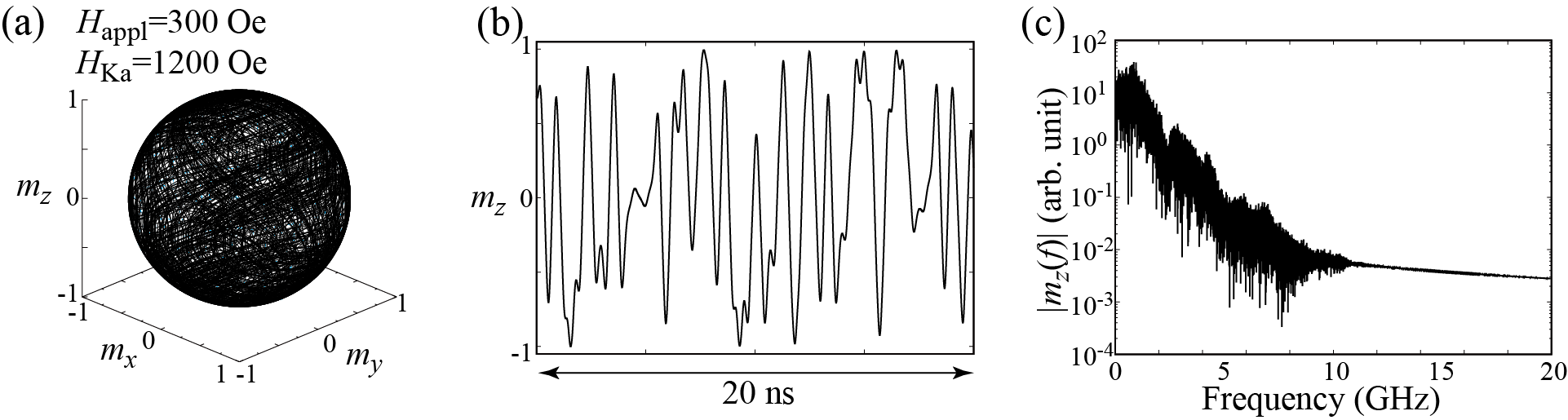}}
\caption{
         (a) Dynamical trajectory of the magnetization in a parametric oscillation state.
              Parameters are $H_{\rm appl}=300$ Oe and $H_{\rm Ka}=1200$ Oe. 
         (b) Time evolution of $m_{z}$ in the parametric oscillation state. 
         (c) Fourier spectrum of $m_{z}$. 
         \vspace{-3ex}}
\label{fig:fig4}
\end{figure}


\subsection*{Chaotic dynamics}
\label{sec:Chaotic dynamics}

In Fig. \ref{fig:fig4}(a), we show the dynamical trajectory of the magnetization for $H_{\rm Ka}=1200$ Oe. 
The dynamical trajectory covers almost the whole region of the unit sphere. 
The time evolution of $m_{z}$ becomes non-periodic, as shown in Fig. \ref{fig:fig4}(b), and the Fourier spectrum shows a broad structure having several peaks. 
These results imply chaotic dynamics of the magnetization \cite{alligood97,ott02}. 
The appearance of chaos for this parameter can also be concluded from the fact that the Lyapunov exponent shown in Fig. \ref{fig:fig3}(b) is positive. 

As mentioned above, chaotic dynamics in the present system are excited because of the presence of the microwave voltage. 
Similar examples in spintronics devices have been found in spin-torque oscillators with time-dependent inputs \cite{li06,yang07,yamaguchi19}. 
The differences of the phenomena observed between the voltage-controlled parametric oscillator studied here and spin-torque oscillator are as follows. 
In the spin-torque oscillators, a sustainable oscillation of the magnetization is driven by direct currents, and an injection of time-dependent inputs is not a necessary condition for the oscillation. 
Spin-torque oscillator cannot show chaotic behavior when only the direct current is injected due to the constraint by the Poincar\'e-Bendixson theorem. 
A way to excite chaos in spin-torque oscillator is to inject time-dependent inputs such as alternating current and/or magnetic field \cite{li06,yang07,yamaguchi19}. 
When the magnitudes of these time-dependent inputs are relatively small, synchronization may occur. 
When their magnitudes are further increased, chaos might be induced. 
On the other hand, the sustainable oscillation of the magnetization in the present voltage-controlled parametric oscillator is driven by microwave voltage. 
In other words, this time-dependent input is a necessary condition for the oscillation. 
Chaotic dynamics appear when the magnitude of the microwave voltage becomes relatively large, as shown in Fig. \ref{fig:fig4}. 

Let us briefly mention applicability of the chaotic dynamics for practical devices. 
Chaos in spintronics devices has been studied both experimentally and theoretically using various methods \cite{kudo06,watelot12,devolder19,bondarenko19,montoya19,williame19,taniguchi19,taniguchi19JMMM,williame20,kamimaki21}. 
An excitation of chaos in spintronics devices may evoke interest from a viewpoint of brain-inspired computing \cite{yamaguchi23,tsunegi23}. 
For example, it was found that a computational capability of physical reservoir computing is enhanced when spin-torque oscillators are near the edge of chaos \cite{yamaguchi23}, where chaos was excited by adding another ferromagnet to the oscillator. 
An excitation of chaos in spin-torque oscillator by time-dependent inputs, however, seems to require large power consumption. 
For example, the amplitude of the alternating current density assumed in the numerical simulations in Refs. \cite{li06,yang07,yamaguchi19} is on the order of $10^{7}-10^{8}$ A/cm${}^{2}$. 
The value is larger than the switching current density of the state of the art spin-transfer torque driven MRAM \cite{dieny16}, and thus, is hardly desirable. 
The large current also causes large power consumption, which is also unsuitable for practical applications. 
The voltage-controlled parametric oscillator, on the other hand, ideally reduces the power consumption significantly. 
In fact, this point has been a motivation for developing VCMA-based MRAM. 
However, these VCMA-based devices often require an external magnetic field for both the switching and parametric oscillation, which is not preferable in practical applications. 
A way to solve the issue might be to use an effective field, instead of applying external magnetic field, such as an interlayer exchange coupling field, as investigated in the study of spin-orbit torque driven MRAM \cite{lau16}. 
Physical reservoir computing by the VCMA-based MRAM without an external magnetic field was investigated recently \cite{taniguchi22srep}, however, switching nor parametric oscillation was used there. 
A development of the voltage-controlled parametric oscillator without requiring an external magnetic field will be of interest as a future work for applying it to practical applications. 



\begin{figure}
\centerline{\includegraphics[width=1.0\columnwidth]{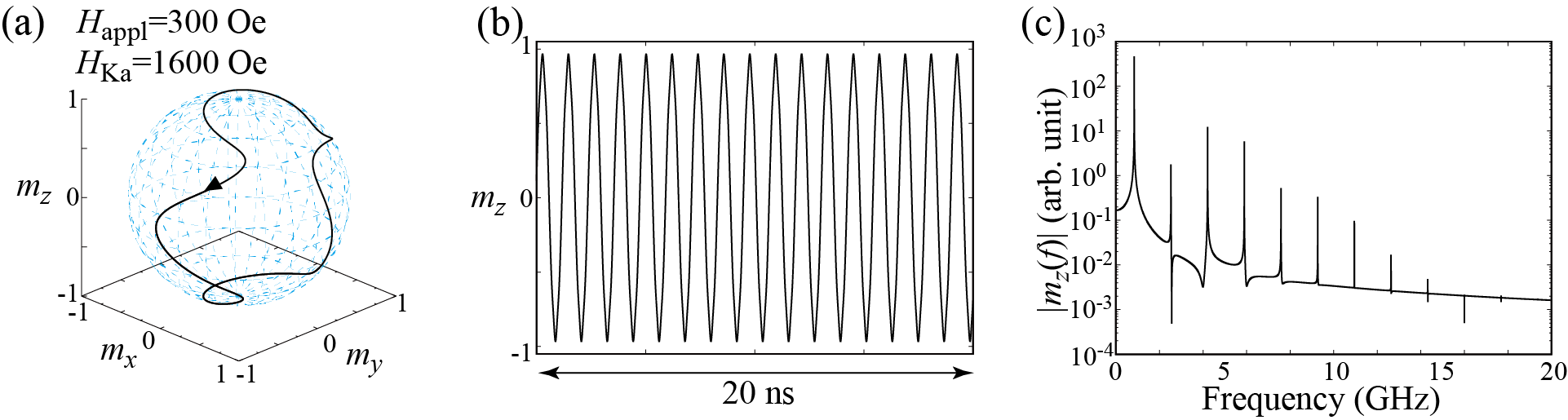}}
\caption{
         (a) Dynamical trajectory of the magnetization in a parametric oscillation state.
              Parameters are $H_{\rm appl}=300$ Oe and $H_{\rm Ka}=1600$ Oe. 
              The black triangle indicates the direction of the magnetization motion. 
         (b) Time evolution of $m_{z}$ in the parametric oscillation state. 
         (c) Fourier spectrum of $m_{z}$. 
         \vspace{-3ex}}
\label{fig:fig5}
\end{figure}


\subsection*{Transient chaos}
\label{sec:Transient chaos}

In Fig. \ref{fig:fig5}(a), we show the dynamical trajectory of the magnetization for $H_{\rm Ka}=1600$ Oe. 
The time evolution of $m_{z}$ and its Fourier transformation are also shown in Figs. \ref{fig:fig5}(b) and \ref{fig:fig5}(c), respectively. 
We note that the dynamics shown here corresponds to $m_{z}(t)$ in a long-time limit, i.e., a steady state. 
The results look similar to those shown in Figs. \ref{fig:fig1} and \ref{fig:fig2}. 
Also, the local maxima of $m_{z}$ in the bifurcation diagram in Fig. \ref{fig:fig3}(a) concentrates on a single point. 
However, there is a critical difference between the magnetization dynamics shown in Fig. \ref{fig:fig5} and those shown in Figs. \ref{fig:fig1} and \ref{fig:fig2}. 


\begin{figure}
\centerline{\includegraphics[width=1.0\columnwidth]{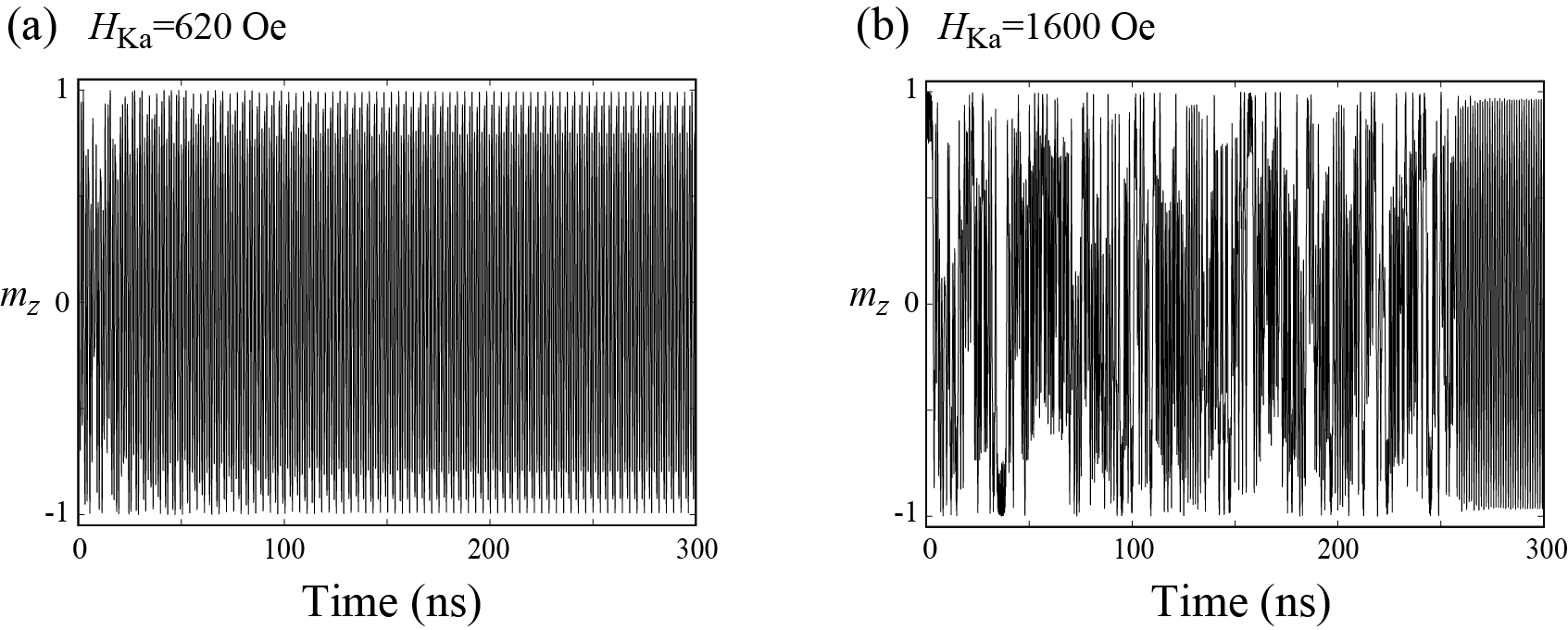}}
\caption{
            Time evolution of $m_{z}$ near $t=0$ for (a) $H_{\rm Ka}=620$ Oe and (b) $H_{\rm Ka}=1600$ Oe. 
         \vspace{-3ex}}
\label{fig:fig6}
\end{figure}



\begin{figure}
\centerline{\includegraphics[width=1.0\columnwidth]{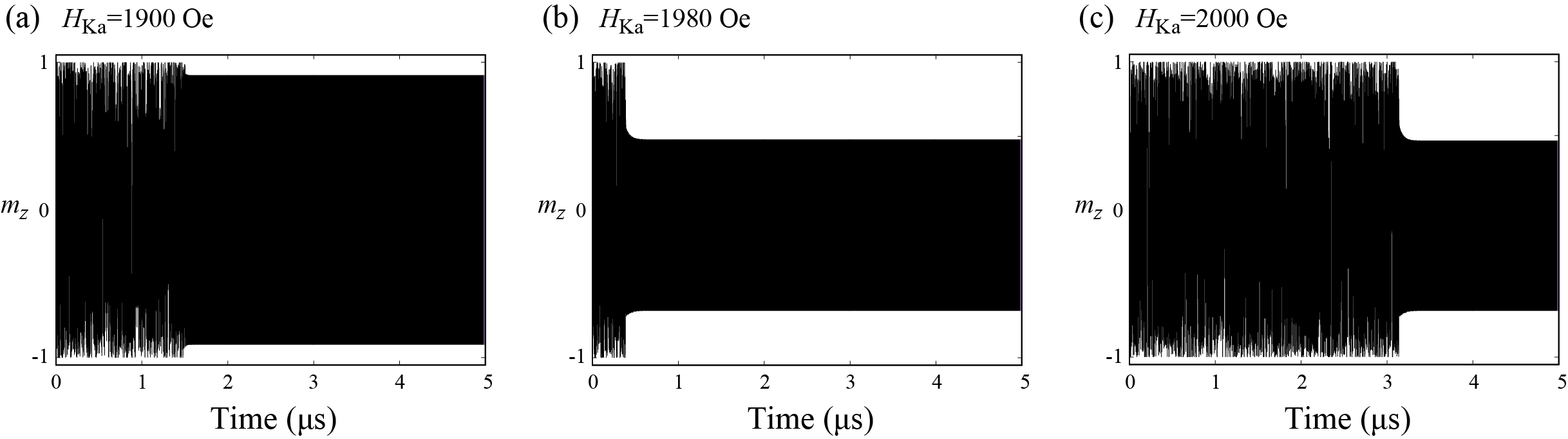}}
\caption{
            Time evolution of $m_{z}$ for (a) $H_{\rm Ka}=1900$ Oe, (b) $H_{\rm Ka}=1980$ Oe, and (c) $H_{\rm Ka}=2000$ Oe. 
         \vspace{-3ex}}
\label{fig:fig7}
\end{figure}


To clarify their differences, it is necessary to focus on a process to reach the steady state. 
In Figs. \ref{fig:fig6}(a) and \ref{fig:fig6}(b), we show the time evolution of $m_{z}(t)$ from $t=0$ to the steady states for $H_{\rm Ka}=620$ Oe and $1600$ Oe, respectively. 
When $H_{\rm Ka}$ is relatively small [see Fig. \ref{fig:fig6}(a)], the magnetization immediately reaches to the steady state shown in Fig. \ref{fig:fig2}(b). 
When $H_{\rm Ka}$ is relatively large [see Fig. \ref{fig:fig6}(b)], on the other hand, chaotic behavior appears initially, and it suddenly changes to the steady state shown in Fig. \ref{fig:fig5}(b). 
The phenomenon shown in Fig. \ref{fig:fig5}(b) corresponds to a transient chaos \cite{lai11}, where dynamical systems initially show chaotic behavior but it suddenly disappears. 
Time necessary to move from chaos to a steady state is sensitive to various parameters and initial conditions of systems \cite{lai11}. 
For example, in Figs. \ref{fig:fig7}(a), \ref{fig:fig7}(b), and \ref{fig:fig7}(c), we show time evolution of $m_{z}$ for $H_{\rm Ka}=1900$ Oe, 1980 Oe, and 2000 Oe, respectively, where the initial conditions are common. 
The results indicate that the time necessary to realize the steady state depends on the parameter $H_{\rm Ka}$ and it does not show, for example, a monotonic change with respect to the change of the parameter. 
The transient chaos in spintronics devices was predicted in a spin-torque oscillator with a delayed-feedback circuit \cite{taniguchi19}. 
The phenomenon has not been verified experimentally yet, although chaos was confirmed recently \cite{kamimaki21}. 


We note that the classification of chaos and transient chaos depends on a measurement time \cite{lai11}. 
For example, in the present work, we solve the LLG equation from $t=0$ to $t=5$ $\mu$s and classify the magnetization dynamics. 
If we change this maximum time ($5$ $\mu$s) to, for example $2$ $\mu$s, the dynamics for $H_{\rm Ka}=2000$ Oe, shown in Fig. \ref{fig:fig7}(c), will be classified as chaos. 
Another example can be seen in the bifurcation diagram and the Lyapunov exponent shown in Fig. \ref{fig:fig3}, where a broadened structure in the bifurcation diagram and a positive Lyapunov for $H_{\rm Ka}=1920$ Oe, indicating chaos for this parameter. 
If we measure the dynamics for this parameter longer, however, the dynamics might change to a steady state; in such a case, the dynamics will be classified as transient chaos. 
As a general knowledge on transient chaos \cite{lai11}, we should remember that the classification of chaos and transient chaos has such an arbitrary property. 



\begin{figure}
\centerline{\includegraphics[width=1.0\columnwidth]{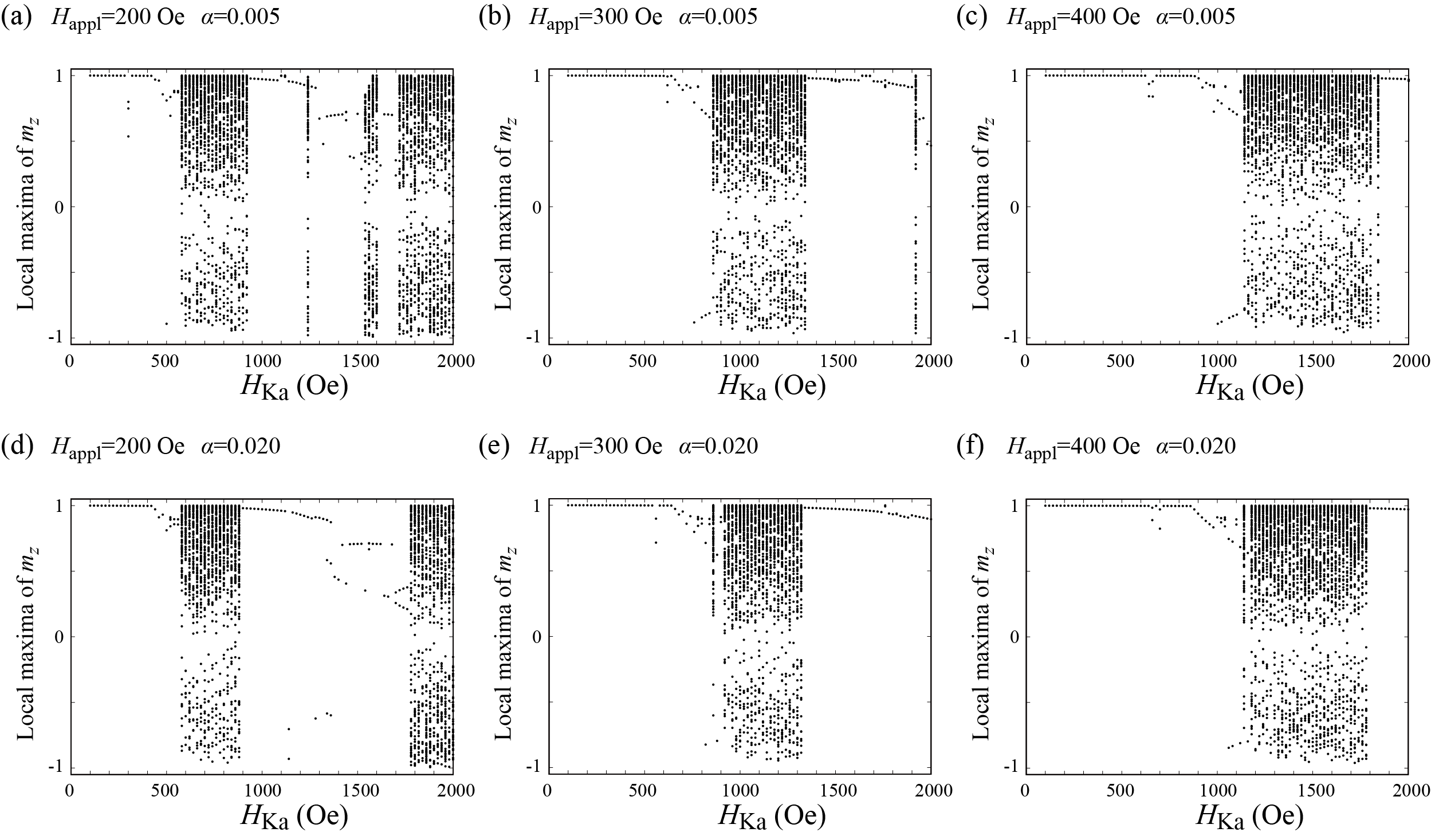}}
\caption{
            Bifurcation diagrams summarizing the local maxima of $m_{z}$ for (a) $H_{\rm appl}=200$ Oe, (b) $H_{\rm appl}=300$ Oe, and (c) $H_{\rm appl}=400$ Oe. 
            Note that (b) is identical to Fig. \ref{fig:fig3}(a) but is shown here for comparison. 
            The value of the damping constant $\alpha$ is $0.005$. 
            Those for $\alpha=0.020$ are shown in (d)-(f). 
         \vspace{-3ex}}
\label{fig:fig8}
\end{figure}


\subsection*{Bifurcation diagrams for various applied fields}
\label{sec:Bifurcation diagrams for various applied fields}

As mentioned above, a bifurcation diagram is useful to classify the magnetization dynamics, although, for example, the difference between a simple steady oscillation and a transient chaos should be verified from temporal dynamics, as mentioned above. 
We also notice that the whole structure of the bifurcation diagram does not change qualitatively even when the initial condition is slightly changed. 
Recall that the present system includes only three parameters, $H_{\rm Ka}$, $H_{\rm appl}$, and $\alpha$. 
Therefore, in Figs. \ref{fig:fig8}(a)-\ref{fig:fig8}(c), we show the bifurcation diagrams for (a) $H_{\rm appl}=200$ Oe, (b) $H_{\rm appl}=300$ Oe, and (c) $H_{\rm appl}=400$ Oe. 
These figures indicate that the local maxima of $m_{z}$ for relatively small $H_{\rm Ka}$ concentrate on single points, which indicate the excitation of the parametric oscillation. 
As $H_{\rm Ka}$ increases, broadened structures appear, i.e., the local maxima of $m_{z}$ have various values, implying the appearance of complex oscillation and chaos. 
The figures also imply that this boundary between the parametric oscillation and the complex dynamics, i.e., the boundary between the single and multiple points in the bifurcation diagram, locates near $H_{\rm Ka}/H_{\rm appl}\simeq 1.5-2.0$. 
We examine similar calculations for different $\alpha$ but the boundary seems to be unchanged; see Figs. \ref{fig:fig8}(d)-\ref{fig:fig8}(f), which are the bifurcation diagrams for $\alpha=0.020$. 
Because of high nonlinearity in the LLG equation, it is difficult to analytically verify these indications; however, it might be useful to design, for example, the modulation voltage for VCMA-based MRAM utilizing the parametric oscillation. 
We keep this question as a future work. 

We note that the value of $H_{\rm Ka}/H_{\rm appl}\simeq 1.5-2.0$ is available by current technology. 
The typical value of $H_{\rm appl}$ is on the order of $100$ Oe; for example, it is $250$ Oe in Ref. \cite{shiota17} and $720$ Oe in Ref. \cite{yamamoto20}. 
Although the value of $H_{\rm appl}$ can be further large experimentally, a large $H_{\rm appl}$ might be unsuitable for practical applications; see Methods for analytical treatment of the LLG equation. 
On the other hand, the value of $H_{\rm Ka}$ relates to the VCMA efficiency $\eta$, the saturation magnetization $M$, the applied voltage $V_{\rm appl}$, and the thicknesses of the free and insulating layers, $d_{\rm F}$ and $d_{\rm I}$, via $H_{\rm Ka}=(2\eta V_{\rm appl})/(Md_{\rm F}d_{\rm I})$. 
Substituting their typical values [$\eta$ is about $300$ fJ/(Vm) \cite{nozaki20}, $V_{\rm appl}$ is about $0.5$ V at maximum, $M$ is about $1000$ emu/cm${}^{3}$, $d_{\rm F}$ is about $1$ nm, and $d_{\rm I}$ is about $2.5$ nm), $H_{\rm Ka}$ can be on the order of kilo Oersted at maximum, as written in the introduction. 
Therefore, we believe that the value of $H_{\rm Ka}/H_{\rm appl}\simeq 1.5-2.0$ is experimentally achievable. 


\section*{Conclusion}
\label{sec:Conclusion}

In summary, the magnetization dynamics in the voltage-controlled parametric oscillator for a large microwave voltage limit were investigated by numerical simulation of the LLG equation. 
As the modulation increases, the dynamical trajectory changes from a simple parametric oscillation to complex oscillations, which are still periodic but have several local maxima in the amplitude. 
Such dynamics will be of little preference in a switching scheme for VCMA-based MRAM. 
The evaluation of the bifurcation diagrams for various values of the external magnetic field and the damping constant indicated that the simple parametric oscillation is broken when the amplitude of the modulated magnetic anisotropy field becomes larger than the external magnetic field, and this bifurcation point seems to be hardly sensitive to the value of the damping constant. 
A further enhancement of the microwave modulation leads to chaotic and transient-chaotic behaviors, which might make the voltage-controlled parametric oscillator applicable to other usage in electric devices. 
These dynamics were classified from the bifurcation diagrams, Lyapunov exponent, and analyses on temporal dynamics. 



\section*{Methods}


\subsection*{Preparation of initial state}
\label{sec:Preparation of initial state}

We prepare natural initial states by solving the LLG equation with thermal fluctuation \cite{taniguchi22}. 
For this purpose, we use Eq. (\ref{eq:field_orig}) as the magnetic field. 
We add a torque, $-\gamma\mathbf{m}\times\mathbf{h}$, due to thermal fluctuation to the right-hand side of Eq. (\ref{eq:LLG}). 
Here, the components of $\mathbf{h}$ satisfy the fluctuation-dissipation theorem \cite{brown63}, 
\begin{equation}
  \langle h_{k}(t) h_{\ell}(t^{\prime}) \rangle 
  =
  \frac{2\alpha k_{\rm B}T}{\gamma MV}
  \delta_{k\ell}
  \delta(t-t^{\prime}). 
\end{equation}
In the numerical simulation, the random field is given by 
\begin{equation}
  h_{k}(t)
  =
  \sqrt{
    \frac{2\alpha k_{\rm B}T}{\gamma MV \Delta t}
  }
  \xi_{k}(t), 
  \label{eq:random_field_numerical}
\end{equation}
where the time increment $\Delta t$ is 1 ps in this work. 
White noise $\xi_{k}$ is defined from two random numbers, $\zeta_{a}$ and $\zeta_{b}$, in the range of $0 < \zeta_{a},\zeta_{b} < 1$ by the Box-Muller transformation as 
$\xi_{a}=\sqrt{-2 \ln \zeta_{a}} \sin(2 \pi \zeta_{b})$ and $\xi_{b}=\sqrt{-2 \ln \zeta_{a}} \cos(2 \pi \zeta_{b})$. 
We added Eq. (\ref{eq:random_field_numerical}) to the magnetic field and solved the LLG equation numerically using the Runge-Kutta method for the investigation of thermal equilibrium before applying microwave voltage. 
The value of the net perpendicular magnetic anisotropy field, $H_{\rm K}-4\pi M (N_{z}-N_{x})$, in the absence of microwave voltage is assumed to be $6.283$ kOe, while the saturation magnetization is set to be $955$ emu/cm${}^{3}$ \cite{yamamoto20}. 
The temperature $T$ is $300$ K, and the volume is $V=\pi \times 50 \times 50 \times 1.1$ nm${}^{3}$, which is typical for VCMA experiments. 
The thermal fluctuation excites a small-amplitude oscillation of the magnetization around the energetically minimum state with the ferromagnetic resonance frequency. 
We pick the temporal directions of the oscillating magnetization and use them as the natural initial states. 
It should be noted that the value of $H_{\rm Ka}$ at which the complex but still periodic oscillation appears weakly depends on the choice of the initial state, despite the structure of the bifurcation diagram is not changed qualitatively. 
It might relates to the presence of two stable phases of the parametric oscillation with respect to the microwave voltage \cite{taniguchi23}. 


\subsection*{Analytical treatment of the LLG equation}
\label{sec:Analytical treatment of the LLG equation}

Here, we discuss an analytical treatment of the parametric oscillation, which provides a condition to excite the oscillation. 
It is, however, not applicable to investigate the bifurcation from the simple parametric oscillation to the complex but still periodic oscillation shown in Figs. \ref{fig:fig1}(b) and \ref{fig:fig2}(a). 

Since we are interested in the oscillation around the $x$ axis, we introduce angles $\Theta$ and $\Phi$ as $\mathbf{m}=(\cos\Theta,\sin\Theta\cos\Phi,\sin\Theta\sin\Phi)$. 
For simplicity, we introduce notations $k=\gamma H_{\rm Ka}$ and $h=\gamma H_{\rm appl}$. 
The LLG equation for $\Theta$ and $\Phi$ are given as $d\Theta/dt = -(1/\sin\Theta)(\partial\varepsilon/\partial\Phi)-\alpha(\partial\varepsilon/\partial\Theta)$ and $\sin\Theta(d\Phi/dt)=(\partial\varepsilon/\partial\Theta)-\alpha(1/\sin\Theta)(\partial\varepsilon/\partial\Phi)$, where 
\begin{equation}
  \varepsilon
  =
  -h
  \cos\Theta
  -
  \frac{k}{2}
  \sin(2\pi ft)
  \sin^{2}\Theta
  \sin^{2}\Phi, 
\end{equation}
i.e., 
\begin{equation}
  \frac{d\Theta}{dt}
  =
  k\sin(2\pi ft)
  \sin\Theta
  \sin\Phi
  \cos\Phi
  -
  \alpha
  \left[
    h
    -
    k\sin(2\pi ft)
    \cos\Theta
    \sin^{2}\Phi
  \right]
  \sin\Theta,
  \label{eq:LLG_theta}
\end{equation}
\begin{equation}
  \frac{d\Phi}{dt}
  =
  h
  -
  k\sin(2\pi ft)
  \cos\Theta
  \sin^{2}\Phi
  +
  \alpha
  k \sin(2\pi ft)
  \sin\Phi
  \cos\Phi . 
  \label{eq:LLG_phi}
\end{equation}
Since we are interested in an oscillation where $m_{y}$ and $m_{z}$ oscillates with the frequency $f_{\rm L}=f/2$, we introduce $\Psi=\Phi-2\pi f_{\rm L}t$, which obeys 
\begin{equation}
\begin{split}
  \frac{d\Psi}{dt}
  =&
  h
  -
  2\pi f_{\rm L}
  -
  k\sin(2\pi ft)
  \cos\Theta
  \sin^{2}(\Psi+2\pi f_{\rm L}t)
  +
  \alpha
  k \sin(2\pi ft)
  \sin(\Psi+2\pi f_{\rm L}t)
  \cos(\Psi+2\pi f_{\rm L}t). 
  \label{eq:LLG_psi}
\end{split}
\end{equation}
In the parametric oscillation states, the tilted angle $\Theta$ of the magnetization from the $x$ axis and the phase difference $\Psi$ are approximately constants \cite{taniguchi23}. 
Therefore, after averaging Eqs. (\ref{eq:LLG_theta}) and (\ref{eq:LLG_psi}) over a period $1/f_{\rm L}$, we obtain, 
\begin{equation}
  \overline{
    \frac{d\Theta}{dt}
  }
  =
  \frac{k}{4}
  \sin\Theta
  \cos 2\Psi
  -
  \alpha
  \left(
    h
    -
    \frac{k}{4}
    \cos\Theta
    \sin 2\Psi
  \right)
  \sin\Theta, 
\end{equation}
\begin{equation}
  \overline{
    \frac{d\Psi}{dt}
  }
  =
  \sigma
  -
  \frac{k}{4}
  \cos\Theta
  \sin 2\Psi
  +
  \frac{\alpha k}{4}
  \cos 2\Psi , 
\end{equation}
where $\sigma=h-2\pi f_{\rm L}$. 
Although the present work focuses on the case of $\sigma=0$ only, we keep $\sigma$ as finite here, for generality. 
The steady state solutions of $\Theta$ and $\Psi$ after averaging are 
\begin{equation}
  \cos\Theta
  =
  \pm
  \frac{4(\sigma + \alpha^{2}h)}{\sqrt{(1+\alpha^{2})k^{2}-[4 \alpha(h-\sigma)]^{2}}}, 
  \label{eq:cos_theta_sol}
\end{equation}
\begin{equation}
  \tan 2\Psi
  =
  \pm
  \frac{\sqrt{(1+\alpha^{2})k^{2}-[4\alpha(h-\sigma)]^{2}}}{4\alpha(h-\sigma)}. 
  \label{eq:tan_psi_sol}
\end{equation}
These solutions imply, for example, that $(1+\alpha^{2})k^{2}>[4\alpha(h-\sigma)]^{2}$ for exciting the parametric oscillation, which can easily be satisfied when $\alpha\ll 1$. 
These solutions, however, cannot be used to discuss, for example, the bifurcation from the simple parametric oscillation to the complex oscillation because $\Phi$ above is assumed to oscillates with only a single frequency $f_{\rm L}$, while the complex oscillation is a superposition of multiple frequencies. 
Even for the parametric oscillation state, the above solutions may not be quantitative due to the fact that, strictly speaking for example, $\Theta$ is not constant. 


Although a reliability of the magnetization switching by the parametric oscillation is not the main scope of this work, let us briefly provide a comment on the relationship between the switching reliability and the value of $H_{\rm Ka}/H_{\rm appl}$ mentioned in the introduction. 
First, the value of $H_{\rm appl}$ should be carefully chosen. 
For a large $H_{\rm appl}$, the precession frequency of the magnetization becomes high. 
Such a fast precession makes the switching unstable because the dynamics becomes sensitive to the pulse shape, as mentioned in the introduction. 
When $H_{\rm appl}$ is small, however, the switching is also unstable because the dynamics is highly affected by thermal fluctuation, which is also written in the introduction. 
Summarizing them, the value of $H_{\rm appl}$ should be carefully determined, depending on the various factors, such as circuit quality controlling the pulse shape and the volume of the ferromagnetic layer. 
Second, a large $H_{\rm Ka}$ is considered to be preferable for a reliable switching due to the following reasons. 
First, as mentioned below Eqs. (\ref{eq:cos_theta_sol}) and (\ref{eq:tan_psi_sol}), there is a threshold value of $H_{\rm Ka}$ [$(1+\alpha^{2})k^{2}>[4\alpha(h-\sigma)]^{2}$] to excite the parametric oscillation. 
Second, Eq. (\ref{eq:cos_theta_sol}) indicates that the cone angle $\Theta$ of the magnetization precession around $H_{\rm appl}$ becomes close to $90^{\circ}$ when $H_{\rm Ka}$ is large. 
In other words, the cone angle decreases as $H_{\rm Ka}$ decreases. 
A precession with a small cone angle is unstable because such a small oscillation is easily disturbed by thermal fluctuation. 
Regarding these points, it will be preferable to increase $H_{\rm Ka}/H_{\rm appl}$, or $H_{\rm Ka}$, for a reliable switching because there is a threshold of $H_{\rm Ka}$ of the parametric oscillation and it is necessary to make the oscillation robust against thermal fluctuation. 
However, as revealed in the main text, the precession becomes complex when $H_{\rm Ka}/H_{\rm appl}$ becomes greatly large, which is a main message in this work. 


\subsection*{Evaluation method of Lyapunov exponent}
\label{sec:Evaluation method of Lyapunov exponent}

The evaluation method of the Lyapunov exponent is as follows \cite{taniguchi22,taniguchi20AIP}. 
We denote the solution of the LLG equation at a certain time $t$ as $\mathbf{m}(t)$. 
We introduce the zenith and azimuth angles, $\theta$ and $\varphi$, as $\mathbf{m}=(m_{x},m_{y},m_{z})=(\sin\theta\cos\varphi,\sin\theta\sin\varphi,\cos\theta)$. 
We also introduce $\mathbf{m}^{(1)}(t)=(\sin\theta^{(1)}\cos\varphi^{(1)},\sin\theta^{(1)}\sin\varphi^{(1)},\cos\theta^{(1)})$, where $\theta^{(1)}$ and $\varphi^{(1)}$ satisfy $\epsilon=\sqrt{[\theta-\theta^{(1)}]^{2}+[\varphi-\varphi^{(1)}]^{2}}$. 
Note that $\epsilon$ corresponds to the distance between $\mathbf{m}(t)$ and $\mathbf{m}^{(1)}(t)$ at $t$ in the spherical space. 
Since the Lyapunov exponent characterizes the sensitivity of the dynamical system to the initial state, we study an expansion of $\epsilon$ with time. 
For this purpose, we assume a small value of $\epsilon$, which is $\epsilon=1.0\times 10^{-5}$ in this work. 
For convenience, we introduce a notation, 
\begin{equation}
  \mathcal{D}[\mathbf{m}(t),\mathbf{m}^{(1)}(t)]
  =
  \sqrt{
    \left[
      \theta(t)
      -
      \theta^{(1)}(t)
    \right]^{2}
    +
    \left[
      \varphi(t)
      -
      \varphi^{(1)}(t)
    \right]^{2}
  }.
\end{equation}
Solving the LLG equations of $\mathbf{m}(t)$ and $\mathbf{m}^{(1)}(t)$, we obtain $\mathbf{m}(t+\Delta t)$ and $\mathbf{m}^{(1)}(t+\Delta t)$, where $\Delta t$ is time increment. 
From them, we evaluate the distance between $\mathbf{m}(t+\Delta t)$ and $\mathbf{m}^{(1)}(t+\Delta t)$ as 
\begin{equation}
  \mathcal{D}[\mathbf{m}(t+\Delta t),\mathbf{m}^{(1)}(t+\Delta t)]
  =
  \sqrt{
    \left[
      \theta(t+\Delta t)
      -
      \theta^{(1)}(t+\Delta t)
    \right]^{2}
    +
    \left[
      \varphi(t+\Delta t)
      -
      \varphi^{(1)}(t+\Delta t)
    \right]^{2}
  }. 
\end{equation}
Then, a temporal Lyapunov exponent at $t+\Delta t$ is given as 
\begin{equation}
  \varLambda^{(1)}
  =
  \frac{1}{\Delta t}
  \ln
  \frac{\mathscr{D}^{(1)}}{\epsilon},
\end{equation}
where $\mathscr{D}^{(1)}=\mathcal{D}[\mathbf{m}(t+\Delta t),\mathbf{m}^{(1)}(t+\Delta t)]$. 

Next, we introduce $\mathbf{m}^{(2)}(t+\Delta t)=(\sin\theta^{(2)}\cos\varphi^{(2)},\sin\theta^{(2)}\sin\varphi^{(2)},\cos\theta^{(2)})$, where $\theta^{(2)}$ and $\varphi^{(2)}$ are defined as 
\begin{align}
&
  \theta^{(2)}(t+\Delta t)
  =
  \theta(t+\Delta t)
  +
  \epsilon
  \frac{\theta^{(1)}(t+\Delta t)-\theta(t+\Delta t)}{\mathcal{D}[\mathbf{m}(t+\Delta t),\mathbf{m}^{(1)}(t+\Delta t)]},
\\
&
  \varphi^{(2)}(t+\Delta t)
  =
  \varphi(t+\Delta t)
  +
  \epsilon
  \frac{\varphi^{(1)}(t+\Delta t)-\varphi(t+\Delta t)}{\mathcal{D}[\mathbf{m}(t+\Delta t),\mathbf{m}^{(1)}(t+\Delta t)]}.
\end{align}
According to these definitions, $\mathbf{m}(t+\Delta t)$ and $\mathbf{m}^{(2)}(t+\Delta t)$ satisfy 
\begin{equation}
  \mathcal{D}[\mathbf{m}(t+\Delta t),\mathbf{m}^{(2)}(t+\Delta t)]
  =
  \epsilon.
\end{equation}
It means that $\mathbf{m}^{(2)}(t+\Delta t)$ is defined by moving $\mathbf{m}(t+\Delta t)$ to the direction of $\mathbf{m}^{(1)}(t+\Delta t)$ with a distance $\epsilon$ in the $(\theta,\varphi)$ phase space. 
Solving the LLG equations for $\mathbf{m}(t+\Delta t)$ and $\mathbf{m}^{(2)}(t+\Delta t)$, we obtain $\mathbf{m}(t+2\Delta t)$ and $\mathbf{m}^{(2)}(t+2\Delta t)$. 
The temporal Lyapunov exponent at $t+2\Delta t$ is 
\begin{equation}
  \varLambda^{(2)}
  =
  \frac{1}{\Delta t}
  \ln
  \frac{\mathscr{D}^{(2)}}{\epsilon},
\end{equation}
where $\mathscr{D}^{(2)}=\mathcal{D}[\mathbf{m}(t+2\Delta t),\mathbf{m}^{(2)}(t+2\Delta t)]$. 

These procedures are generalized. 
At $t+n\Delta t$, we have $\mathbf{m}(t+n\Delta t)=(\sin\theta(t+n\Delta t)\cos\varphi(t+n\Delta),\sin\theta(t+n\Delta t)\sin\varphi(t+n\Delta t),\cos\theta(t+n\Delta t))$ and $\mathbf{m}^{(n)}(t+n\Delta t)=(\sin\theta^{(n)}(t+n\Delta t)\cos\varphi^{(n)}(t+n\Delta),\sin\theta^{(n)}(t+n\Delta t)\sin\varphi^{(n)}(t+n\Delta t),\cos\theta^{(n)}(t+n\Delta t))$. 
Then, we define $\mathbf{m}^{(n+1)}(t+n\Delta t)=(\sin\theta^{(n+1)}(t+n\Delta t)\cos\varphi^{(n+1)}(t+n\Delta),\sin\theta^{(n+1)}(t+n\Delta t)\sin\varphi^{(n+1)}(t+n\Delta t),\cos\theta^{(n+1)}(t+n\Delta t))$ by moving $\mathbf{m}(t+n\Delta t)$ to the direction of $\mathbf{m}^{(n)}(t+n\Delta t)$ with a distance $\epsilon$ in the phase space as 
\begin{align}
&
  \theta^{(n+1)}(t+n\Delta t)
  =
  \theta(t+n\Delta t)
  +
  \epsilon
  \frac{\theta^{(n)}(t+n\Delta t)-\theta(t+n\Delta t)}{\mathcal{D}[\mathbf{m}(t+n\Delta t),\mathbf{m}^{(n)}(t+n\Delta t)]},
\\
&
  \varphi^{(n+1)}(t+n\Delta t)
  =
  \varphi(t+n\Delta t)
  +
  \epsilon
  \frac{\varphi^{(n)}(t+n\Delta t)-\varphi(t+n\Delta t)}{\mathcal{D}[\mathbf{m}(t+n\Delta t),\mathbf{m}^{(n)}(t+n\Delta t)]}. 
\end{align}
Note that $\mathbf{m}(t+n\Delta t)$ and $\mathbf{m}^{(n+1)}(t+n\Delta t)$ satisfy $\mathcal{D}[\mathbf{m}(t+n\Delta t),\mathbf{m}^{(n+1)}(t+n\Delta t)] =\epsilon$. 
Then, solving the LLG equations of $\mathbf{m}(t+n\Delta t)$ and $\mathbf{m}^{(n+1)}(t+n\Delta t)$, we obtain $\mathbf{m}(t+(n+1)\Delta t)$ and $\mathbf{m}^{(n+1)}(t+(n+1)\Delta t)$. 
A temporal Lyapunov exponent at $t+(n+1)\Delta t$ is 
\begin{equation}
  \varLambda^{(n+1)}
  =
  \frac{1}{\Delta t}
  \ln
  \frac{\mathscr{D}^{(n+1)}}{\epsilon},
\end{equation}
where $\mathscr{D}^{(n+1)}=\mathcal{D}[\mathbf{m}(t+(n+1)\Delta t),\mathbf{m}^{(n+1)}(t+(n+1)\Delta t)]$. 
The Lyapunov exponent is defined as a long-time average of the temporal Lyapunov exponent as 
\begin{equation}
  \varLambda
  =
  \lim_{N \to \infty}
  \frac{1}{N}
  \sum_{i=1}^{N}
  \varLambda^{(i)}. 
\end{equation}
In the present study, we solve the LLG equation from $t=0$ to $t_{\rm max}=5$ $\mu$s with $\Delta t=1$ ps. 
Therefore, there are $t_{\rm max}/\Delta t=5 \times 10^{6}$ steps during the evaluation of the magnetization dynamics. 
We use the last $1 \times 10^{6}$ steps for the evaluation of the Lyapunov exponent. 
Note that this method is an application of Shimada-Nagashima method \cite{shimada79} for the evaluation of the Lyapunov exponent from numerical simulation of an equation of motion to the LLG equation. 
Strictly speaking, the Lyapunov exponent evaluated here corresponds to the conditional largest Lyapunov exponent in non-autonomous system \cite{taniguchi22,yamaguchi19}.

\section*{Acknowledgements}

The authors are grateful to Takayuki Nozaki for discussion. 
This work was supported by a JSPS KAKENHI Grant, Number 20H05655.


\section*{Author contributions statement}

T.T. designed the project, performed the numerical simulations, prepared figures, and wrote the manuscript.  


\section*{Competing interests}

The authors declare no competing interests. 


\section*{Data availability}

The datasets used and/or analyses during the current study available from the corresponding author on reasonable request.


\section*{Additional information}

\textbf{Correspondence} and requests for materials should be addressed to T.T.






\begin{thebibliography}{10}
\urlstyle{rm}
\expandafter\ifx\csname url\endcsname\relax
  \def\url#1{\texttt{#1}}\fi
\expandafter\ifx\csname urlprefix\endcsname\relax\def\urlprefix{URL }\fi
\expandafter\ifx\csname doiprefix\endcsname\relax\def\doiprefix{DOI: }\fi
\providecommand{\bibinfo}[2]{#2}
\providecommand{\eprint}[2][]{\url{#2}}

\bibitem{weisheit07}
\bibinfo{author}{Weisheit, M.} \emph{et~al.}
\newblock \bibinfo{journal}{\bibinfo{title}{Electric {Field}-{Induced}
  {Modification} of {Magnetism} in {Thin}-{Film} {Ferromagnets}}}.
\newblock {\emph{\JournalTitle{Science}}} \textbf{\bibinfo{volume}{315}},
  \bibinfo{pages}{349} (\bibinfo{year}{2007}).

\bibitem{duan08}
\bibinfo{author}{Duan, C.-G.} \emph{et~al.}
\newblock \bibinfo{journal}{\bibinfo{title}{Surface {Magnetoelectric} {Effect}
  in {Ferromagnetic} {Metal} {Films}}}.
\newblock {\emph{\JournalTitle{Phys. Rev. Lett.}}}
  \textbf{\bibinfo{volume}{101}}, \bibinfo{pages}{137201}
  (\bibinfo{year}{2008}).

\bibitem{nakamura09}
\bibinfo{author}{Nakamura, K.} \emph{et~al.}
\newblock \bibinfo{journal}{\bibinfo{title}{Giant {Modification} of the
  {Magnetocrystalline} {Anisotropy} in {Transition}-{Metal} {Monolayers} by an
  {External} {Electric} {Field}}}.
\newblock {\emph{\JournalTitle{Phys. Rev. Lett.}}}
  \textbf{\bibinfo{volume}{102}}, \bibinfo{pages}{187201}
  (\bibinfo{year}{2009}).

\bibitem{tsujikawa09}
\bibinfo{author}{Tsujikawa, M.} \& \bibinfo{author}{Oda, T.}
\newblock \bibinfo{journal}{\bibinfo{title}{Finite {Electric} {Field} {Effects}
  in the {Large} {Perpendicular} {Magnetic} {Anisotropy} {Surface}
  {Pt}/{Fe}/{Pt}(001): {A} {First}-{Principles} {Study}}}.
\newblock {\emph{\JournalTitle{Phys. Rev. Lett.}}}
  \textbf{\bibinfo{volume}{102}}, \bibinfo{pages}{247203}
  (\bibinfo{year}{2009}).

\bibitem{miwa17}
\bibinfo{author}{Miwa, S.} \emph{et~al.}
\newblock \bibinfo{journal}{\bibinfo{title}{Voltage controlled interfacial
  magnetism through platinum orbits}}.
\newblock {\emph{\JournalTitle{Nat. Commun.}}} \textbf{\bibinfo{volume}{8}},
  \bibinfo{pages}{15848} (\bibinfo{year}{2017}).

\bibitem{slonczewski96}
\bibinfo{author}{Slonczewski, J.~C.}
\newblock \bibinfo{journal}{\bibinfo{title}{Current-driven excitation of
  magnetic multilayers}}.
\newblock {\emph{\JournalTitle{J. Magn. Magn. Mater.}}}
  \textbf{\bibinfo{volume}{159}}, \bibinfo{pages}{L1} (\bibinfo{year}{1996}).

\bibitem{berger96}
\bibinfo{author}{Berger, L.}
\newblock \bibinfo{journal}{\bibinfo{title}{Emission of spin waves by a
  magnetic multilayer traversed by a current}}.
\newblock {\emph{\JournalTitle{Phys. Rev. B}}} \textbf{\bibinfo{volume}{54}},
  \bibinfo{pages}{9353} (\bibinfo{year}{1996}).

\bibitem{maruyama09}
\bibinfo{author}{Maruyama, T.} \emph{et~al.}
\newblock \bibinfo{journal}{\bibinfo{title}{Large voltage-induced magnetic
  anisotropy change in a few atomic layers of iron}}.
\newblock {\emph{\JournalTitle{Nat. Nanotechnol.}}}
  \textbf{\bibinfo{volume}{4}}, \bibinfo{pages}{158} (\bibinfo{year}{2009}).

\bibitem{shiota09}
\bibinfo{author}{Shiota, Y.} \emph{et~al.}
\newblock \bibinfo{journal}{\bibinfo{title}{Voltage-{Assisted} {Magnetization}
  {Switching} in {Ultrahin} {Fe}${}_{80}$ {Co}${}_{20}$ {Alloy} {Layers}}}.
\newblock {\emph{\JournalTitle{Appl. Phys. Express}}}
  \textbf{\bibinfo{volume}{2}}, \bibinfo{pages}{063001} (\bibinfo{year}{2009}).

\bibitem{nozaki10}
\bibinfo{author}{Nozaki, T.}, \bibinfo{author}{Shiota, Y.},
  \bibinfo{author}{Shiraishi, M.}, \bibinfo{author}{Shinjo, T.} \&
  \bibinfo{author}{Suzuki, Y.}
\newblock \bibinfo{journal}{\bibinfo{title}{Voltage-induced perpendicular
  magnetic anisotropy change in magnetic tunnel junctions}}.
\newblock {\emph{\JournalTitle{Appl. Phys. Lett.}}}
  \textbf{\bibinfo{volume}{96}}, \bibinfo{pages}{022506}
  (\bibinfo{year}{2010}).

\bibitem{endo10}
\bibinfo{author}{Endo, M.}, \bibinfo{author}{Kanai, S.},
  \bibinfo{author}{Ikeda, S.}, \bibinfo{author}{Matsukura, F.} \&
  \bibinfo{author}{Ohno, H.}
\newblock \bibinfo{journal}{\bibinfo{title}{Electric-field effects on thickness
  dependent magnetic anisotropy of sputtered
  {Mg}{O}/{Co}${}_{40}${Fe}${}_{40}${B}${}_{20}$/{Ta} structures}}.
\newblock {\emph{\JournalTitle{Appl. Phys. Lett.}}}
  \textbf{\bibinfo{volume}{96}}, \bibinfo{pages}{212503}
  (\bibinfo{year}{2010}).

\bibitem{wang11}
\bibinfo{author}{Wang, W.-G.}, \bibinfo{author}{Li, M.},
  \bibinfo{author}{Hageman, S.} \& \bibinfo{author}{Chien, C.~L.}
\newblock \bibinfo{journal}{\bibinfo{title}{Electric-field-assisted switching
  in magnetic tunnel junctions}}.
\newblock {\emph{\JournalTitle{Nat. Mater.}}} \textbf{\bibinfo{volume}{11}},
  \bibinfo{pages}{64} (\bibinfo{year}{2011}).

\bibitem{nozaki13}
\bibinfo{author}{Nozaki, T.} \emph{et~al.}
\newblock \bibinfo{journal}{\bibinfo{title}{Voltage-{Induced} {Magnetic}
  {Anisotropy} {Changes} in an {Ultrathin} {Fe}{B} {Layer} {Sandwiched} between
  {Two} {Mg}{O} {Layers}}}.
\newblock {\emph{\JournalTitle{Appl. Phys. Express}}}
  \textbf{\bibinfo{volume}{6}}, \bibinfo{pages}{073005} (\bibinfo{year}{2013}).

\bibitem{okada14}
\bibinfo{author}{Okada, A.} \emph{et~al.}
\newblock \bibinfo{journal}{\bibinfo{title}{Electric-field effects on magnetic
  anisotropy and damping constant in {Ta}/{Co}{Fe}{B}/{Mg}{O} investigated by
  ferromagnetic resonance}}.
\newblock {\emph{\JournalTitle{Appl. Phys. Lett.}}}
  \textbf{\bibinfo{volume}{105}}, \bibinfo{pages}{052415}
  (\bibinfo{year}{2014}).

\bibitem{skowronski15}
\bibinfo{author}{Skowro\'nski, W.} \emph{et~al.}
\newblock \bibinfo{journal}{\bibinfo{title}{Perpendicular magnetic anisotropy
  of {Ir}/{Co}{Fe}{B}/{Mg}{O} trilayer system tuned by electric fields}}.
\newblock {\emph{\JournalTitle{Appl. Phys. Express}}}
  \textbf{\bibinfo{volume}{8}}, \bibinfo{pages}{053003} (\bibinfo{year}{2015}).

\bibitem{nozaki17}
\bibinfo{author}{Nozaki, T.} \emph{et~al.}
\newblock \bibinfo{journal}{\bibinfo{title}{Highly effcient voltage control of
  spin and enhanced interfacial perpendicular magnetic anisotropy in
  iridium-doped {Fe}/{Mg}{O} magnetic tunnel junctions}}.
\newblock {\emph{\JournalTitle{NPG Asia Mater.}}} \textbf{\bibinfo{volume}{9}},
  \bibinfo{pages}{e451} (\bibinfo{year}{2017}).

\bibitem{nozaki18}
\bibinfo{author}{Nozaki, T.} \emph{et~al.}
\newblock \bibinfo{journal}{\bibinfo{title}{Enhancement in the interfacial
  perpendicular magnetic anisotropy and the voltage-controlled magnetic
  anisotropy by heavy metal doping at the {Fe}/{Mg}{O} interface}}.
\newblock {\emph{\JournalTitle{APL Mater.}}} \textbf{\bibinfo{volume}{6}},
  \bibinfo{pages}{026101} (\bibinfo{year}{2018}).

\bibitem{nozaki20}
\bibinfo{author}{Nozaki, T.} \emph{et~al.}
\newblock \bibinfo{journal}{\bibinfo{title}{Voltage-cotrolled magnetic
  anisotropy in an ultrathin {Ir}-doped {Fe} layer with a {Co}{Fe} termination
  layer}}.
\newblock {\emph{\JournalTitle{APL Mater.}}} \textbf{\bibinfo{volume}{8}},
  \bibinfo{pages}{011108} (\bibinfo{year}{2020}).

\bibitem{shiota12}
\bibinfo{author}{Shiota, Y.} \emph{et~al.}
\newblock \bibinfo{journal}{\bibinfo{title}{Pulse voltage-induced dynamic
  magnetization switching in magnetic tunnel junctions with high
  resistance-area product}}.
\newblock {\emph{\JournalTitle{Appl. Phys. Lett.}}}
  \textbf{\bibinfo{volume}{101}}, \bibinfo{pages}{102406}
  (\bibinfo{year}{2012}).

\bibitem{kanai12}
\bibinfo{author}{Kanai, S.} \emph{et~al.}
\newblock \bibinfo{journal}{\bibinfo{title}{Electric field-induced
  magnetization reversal in a perpendicular-anisotropy {Co}{Fe}{B}-{Mg}{O}
  magnetic tunnel junction}}.
\newblock {\emph{\JournalTitle{Appl. Phys. Lett.}}}
  \textbf{\bibinfo{volume}{101}}, \bibinfo{pages}{122403}
  (\bibinfo{year}{2012}).

\bibitem{amiri13}
\bibinfo{author}{Amiri, P.}, \bibinfo{author}{Upadhyaya, P.},
  \bibinfo{author}{Alzate, J.~G.} \& \bibinfo{author}{Wang, K.~L.}
\newblock \bibinfo{journal}{\bibinfo{title}{Electric-field-induced thermally
  assisted switching of monodomain magnetic bits}}.
\newblock {\emph{\JournalTitle{J. Appl. Phys.}}}
  \textbf{\bibinfo{volume}{113}}, \bibinfo{pages}{013912}
  (\bibinfo{year}{2013}).

\bibitem{grezes16}
\bibinfo{author}{Grezes, C.} \emph{et~al.}
\newblock \bibinfo{journal}{\bibinfo{title}{Ultra-low switching energy and
  scaling in electric-field-controlled nanoscale magnetic tunnel junctions with
  high resistance-area product}}.
\newblock {\emph{\JournalTitle{Appl. Phys. Lett.}}}
  \textbf{\bibinfo{volume}{108}}, \bibinfo{pages}{012403}
  (\bibinfo{year}{2016}).

\bibitem{shiota17}
\bibinfo{author}{Shiota, Y.} \emph{et~al.}
\newblock \bibinfo{journal}{\bibinfo{title}{Reduction in write error rate of
  voltage-driven dynamic magnetization switching by improving thermal stability
  factor}}.
\newblock {\emph{\JournalTitle{Appl. Phys. Lett.}}}
  \textbf{\bibinfo{volume}{111}}, \bibinfo{pages}{022408}
  (\bibinfo{year}{2017}).

\bibitem{lee17}
\bibinfo{author}{Lee, H.} \emph{et~al.}
\newblock \bibinfo{journal}{\bibinfo{title}{A word line pulse circuit technique
  for reliable magnetoelectric random access memory}}.
\newblock {\emph{\JournalTitle{IEEE Trans. Very Large Scle Integ. Syst.}}}
  \textbf{\bibinfo{volume}{7}}, \bibinfo{pages}{3302} (\bibinfo{year}{2017}).

\bibitem{deng17}
\bibinfo{author}{Deng, J.}, \bibinfo{author}{Liang, G.} \&
  \bibinfo{author}{Gupta, G.}
\newblock \bibinfo{journal}{\bibinfo{title}{Ultrafast and lo-energy switching
  in voltage-controlled elliptical p{M}{T}{J}}}.
\newblock {\emph{\JournalTitle{Sci. Rep.}}} \textbf{\bibinfo{volume}{25}},
  \bibinfo{pages}{2027} (\bibinfo{year}{2017}).

\bibitem{song18}
\bibinfo{author}{Song, J.} \emph{et~al.}
\newblock \bibinfo{journal}{\bibinfo{title}{Evaluation of operating margin and
  switching probability of voltage-controlled magnetic anisotropy magnetic
  tunnel junctions}}.
\newblock {\emph{\JournalTitle{IEEE J. Explor. Solid-State Comput. Devices
  Circuits}}} \textbf{\bibinfo{volume}{4}}, \bibinfo{pages}{76}
  (\bibinfo{year}{2018}).

\bibitem{wu20}
\bibinfo{author}{Wu, Y.~C.} \emph{et~al.}
\newblock \bibinfo{journal}{\bibinfo{title}{Study of precessional switching
  speed control in voltage-controlled perpendicular magnetic tunnel junction}}.
\newblock {\emph{\JournalTitle{AIP Adv.}}} \textbf{\bibinfo{volume}{10}},
  \bibinfo{pages}{035123} (\bibinfo{year}{2020}).

\bibitem{shao22}
\bibinfo{author}{Shao, Y.} \emph{et~al.}
\newblock \bibinfo{journal}{\bibinfo{title}{Sub-volt switching of nanoscale
  voltage-controlled perpendicular magnetic tunnel junctions}}.
\newblock {\emph{\JournalTitle{Commun. Mater.}}} \textbf{\bibinfo{volume}{3}},
  \bibinfo{pages}{87} (\bibinfo{year}{2022}).

\bibitem{yamamoto20}
\bibinfo{author}{Yamamoto, T.} \emph{et~al.}
\newblock \bibinfo{journal}{\bibinfo{title}{Voltage-{Driven} {Magnetization}
  {Switching} {Controlled} by {Microwave} {Electric} {Field} {Pumping}}}.
\newblock {\emph{\JournalTitle{Nano Lett.}}} \textbf{\bibinfo{volume}{20}},
  \bibinfo{pages}{6012} (\bibinfo{year}{2020}).

\bibitem{yakata09}
\bibinfo{author}{Yakata, S.} \emph{et~al.}
\newblock \bibinfo{journal}{\bibinfo{title}{Influnence of perpendicular
  magnetic anisotropy on spin-transfer switching current in
  {C}o{F}e{B}/{M}g{O}/{C}o{F}e{B} magnetic tunnel junctions}}.
\newblock {\emph{\JournalTitle{J. Appl. Phys.}}}
  \textbf{\bibinfo{volume}{105}}, \bibinfo{pages}{07D131}
  (\bibinfo{year}{2009}).

\bibitem{ikeda10}
\bibinfo{author}{Ikeda, S.} \emph{et~al.}
\newblock \bibinfo{journal}{\bibinfo{title}{A perpendicular-anisotropy
  {C}o{F}e{B}-{M}g{O} magnetic tunnel junction}}.
\newblock {\emph{\JournalTitle{Nat. Mater.}}} \textbf{\bibinfo{volume}{9}},
  \bibinfo{pages}{721} (\bibinfo{year}{2010}).

\bibitem{kubota12}
\bibinfo{author}{Kubota, H.} \emph{et~al.}
\newblock \bibinfo{journal}{\bibinfo{title}{Enhancement of perpendicular
  magnetic anisotropy in {F}e{B} free layers using a thin {M}g{O} cap layer}}.
\newblock {\emph{\JournalTitle{J. Appl. Phys.}}}
  \textbf{\bibinfo{volume}{111}}, \bibinfo{pages}{07C723}
  (\bibinfo{year}{2012}).

\bibitem{shiota16}
\bibinfo{author}{Shiota, Y.} \emph{et~al.}
\newblock \bibinfo{journal}{\bibinfo{title}{Evaluation of write error rate for
  voltage-driven dynamic magnetization switching in magnetic tunnel junctions
  with perpendicular magnetization}}.
\newblock {\emph{\JournalTitle{Appl. Phys. Express}}}
  \textbf{\bibinfo{volume}{9}}, \bibinfo{pages}{013001} (\bibinfo{year}{2016}).

\bibitem{taniguchi22}
\bibinfo{author}{Taniguchi, T.}
\newblock \bibinfo{journal}{\bibinfo{title}{Non-periodic input-driven
  magnetization dynamics in voltage-controlled parametric oscillator}}.
\newblock {\emph{\JournalTitle{J. Magn. Magn. Mater.}}}
  \textbf{\bibinfo{volume}{563}}, \bibinfo{pages}{170009}
  (\bibinfo{year}{2022}).

\bibitem{taniguchi23}
\bibinfo{author}{Taniguchi, T.}
\newblock \bibinfo{journal}{\bibinfo{title}{Phase locking in voltage-controlled
  parametric oscillator}}.
\newblock {\emph{\JournalTitle{J. Magn. Magn. Mater.}}}
  \textbf{\bibinfo{volume}{578}}, \bibinfo{pages}{170806}
  (\bibinfo{year}{2023}).

\bibitem{strogatz01}
\bibinfo{author}{Strogatz, S.~H.}
\newblock \emph{\bibinfo{title}{Nonlinear Dynamics and Chaos: With Applications
  to Physics, Biology, Chemistry, and Engineering}}
  (\bibinfo{publisher}{Westview Press}, \bibinfo{year}{2001}),
  \bibinfo{edition}{first} edn.

\bibitem{verba14}
\bibinfo{author}{Verba, R.}, \bibinfo{author}{Tiberkevich, V.},
  \bibinfo{author}{Krivorotov, I.} \& \bibinfo{author}{Slavin, A.}
\newblock \bibinfo{journal}{\bibinfo{title}{Parametric {Excitation} of {Spin}
  {Waves} by {Voltage}-{Controlled} {Magnetic} {Anisotropy}}}.
\newblock {\emph{\JournalTitle{Phys. Rev. Applied}}}
  \textbf{\bibinfo{volume}{1}}, \bibinfo{pages}{044006} (\bibinfo{year}{2014}).

\bibitem{verba16}
\bibinfo{author}{Verba, R.}, \bibinfo{author}{Finocchio, M. C.~G.},
  \bibinfo{author}{Tiberkevich, V.} \& \bibinfo{author}{Slavin, A.}
\newblock \bibinfo{journal}{\bibinfo{title}{Excitation of propagating spin
  waves in ferromagnetic nanowires by microwave voltage-controlled magnetic
  anisotropy}}.
\newblock {\emph{\JournalTitle{Sci. Rep.}}} \textbf{\bibinfo{volume}{6}},
  \bibinfo{pages}{25018} (\bibinfo{year}{2016}).

\bibitem{chen17}
\bibinfo{author}{Chen, Y.-J.} \emph{et~al.}
\newblock \bibinfo{journal}{\bibinfo{title}{Parametric {Resonance} of
  {Magnetization} {Excited} by {Electric} {Field}}}.
\newblock {\emph{\JournalTitle{Nano Lett.}}} \textbf{\bibinfo{volume}{17}},
  \bibinfo{pages}{572} (\bibinfo{year}{2017}).

\bibitem{rana17}
\bibinfo{author}{Rana, B.}, \bibinfo{author}{Fukuma, Y.},
  \bibinfo{author}{Miura, K.}, \bibinfo{author}{Takahashi, H.} \&
  \bibinfo{author}{Otani, Y.}
\newblock \bibinfo{journal}{\bibinfo{title}{Excitation of coherent propagating
  spin waves in ultrathin {Co}{Fe}{B} film by voltage-controlled magnetic
  anisotropy}}.
\newblock {\emph{\JournalTitle{Appl. Phys. Lett.}}}
  \textbf{\bibinfo{volume}{111}}, \bibinfo{pages}{052404}
  (\bibinfo{year}{2017}).

\bibitem{alligood97}
\bibinfo{author}{Alligood, K.~T.}, \bibinfo{author}{Sauer, T.~D.} \&
  \bibinfo{author}{Yorke, J.~A.}
\newblock \emph{\bibinfo{title}{Chaos. An Introduction to Dynamical Systems}}
  (\bibinfo{publisher}{Spinger (New York)}, \bibinfo{year}{1997}).

\bibitem{ott02}
\bibinfo{author}{Ott, E.}
\newblock \emph{\bibinfo{title}{Chaos in Dynamical Systems}}
  (\bibinfo{publisher}{Cambridge University Press (Cambridge)},
  \bibinfo{year}{2002}), \bibinfo{edition}{second} edn.

\bibitem{lai11}
\bibinfo{author}{Lai, Y.-C.} \& \bibinfo{author}{T\'el, T.}
\newblock \emph{\bibinfo{title}{Transient Chaos}} (\bibinfo{publisher}{Spinger
  (New York)}, \bibinfo{year}{2011}).

\bibitem{shimada79}
\bibinfo{author}{Shimada, I.} \& \bibinfo{author}{Nagashima, T.}
\newblock \bibinfo{journal}{\bibinfo{title}{A {Numerical} {Approach} to
  {Ergodic} {Problem} of {Dissipative} {Dynamical} {Systems}}}.
\newblock {\emph{\JournalTitle{Prog. Theor. Phys.}}}
  \textbf{\bibinfo{volume}{61}}, \bibinfo{pages}{1605} (\bibinfo{year}{1979}).

\bibitem{li06}
\bibinfo{author}{Li, Z.}, \bibinfo{author}{Li, Y.~C.} \&
  \bibinfo{author}{Zhang, S.}
\newblock \bibinfo{journal}{\bibinfo{title}{Dynamic magnetization states of a
  spin valve in the presence of dc and ac currents: {Synchronization},
  modification, and chaos}}.
\newblock {\emph{\JournalTitle{Phys. Rev. B}}} \textbf{\bibinfo{volume}{74}},
  \bibinfo{pages}{054417} (\bibinfo{year}{2006}).

\bibitem{yang07}
\bibinfo{author}{Yang, Z.}, \bibinfo{author}{Zhang, S.} \& \bibinfo{author}{Li,
  Y.~C.}
\newblock \bibinfo{journal}{\bibinfo{title}{Chaotic {Dynamics} of
  {Spin}-{Valve} {Oscillators}}}.
\newblock {\emph{\JournalTitle{Phys. Rev. Lett.}}}
  \textbf{\bibinfo{volume}{99}}, \bibinfo{pages}{134101}
  (\bibinfo{year}{2007}).

\bibitem{yamaguchi19}
\bibinfo{author}{Yamaguchi, T.} \emph{et~al.}
\newblock \bibinfo{journal}{\bibinfo{title}{Synchronization and chaos in a
  spin-torque oscillator with a perpendicularly magnetized free layer}}.
\newblock {\emph{\JournalTitle{Phys. Rev. B}}} \textbf{\bibinfo{volume}{100}},
  \bibinfo{pages}{224422} (\bibinfo{year}{2019}).

\bibitem{kudo06}
\bibinfo{author}{Kudo, K.}, \bibinfo{author}{Sato, R.} \&
  \bibinfo{author}{Mizushima, K.}
\newblock \bibinfo{journal}{\bibinfo{title}{Synchronized {Magnetization}
  {Oscillations} in {F}/{N}/{F} {Nanopillars}}}.
\newblock {\emph{\JournalTitle{Jpn. J. Appl. Phys.}}}
  \textbf{\bibinfo{volume}{45}}, \bibinfo{pages}{3869} (\bibinfo{year}{2006}).

\bibitem{watelot12}
\bibinfo{author}{Petit-Watelot, S.} \emph{et~al.}
\newblock \bibinfo{journal}{\bibinfo{title}{Commensurability and chaos in
  magnetic vortex oscillations}}.
\newblock {\emph{\JournalTitle{Nat. Phys.}}} \textbf{\bibinfo{volume}{8}},
  \bibinfo{pages}{682--687} (\bibinfo{year}{2012}).

\bibitem{devolder19}
\bibinfo{author}{Devolder, T.} \emph{et~al.}
\newblock \bibinfo{journal}{\bibinfo{title}{Chaos in magnetic nanocontact
  vortex oscillators}}.
\newblock {\emph{\JournalTitle{Phys. Rev. Lett.}}}
  \textbf{\bibinfo{volume}{123}}, \bibinfo{pages}{147701}
  (\bibinfo{year}{2019}).

\bibitem{bondarenko19}
\bibinfo{author}{Bondarenko, A.~V.}, \bibinfo{author}{Holmgren, E.},
  \bibinfo{author}{Li, Z.~W.}, \bibinfo{author}{Ivanov, B.~A.} \&
  \bibinfo{author}{Korenivski, V.}
\newblock \bibinfo{journal}{\bibinfo{title}{Chaotic dynamics in spin-vortex
  pairs}}.
\newblock {\emph{\JournalTitle{Phys. Rev. B}}} \textbf{\bibinfo{volume}{99}},
  \bibinfo{pages}{054402} (\bibinfo{year}{2019}).

\bibitem{montoya19}
\bibinfo{author}{Montoya, E.~A.} \emph{et~al.}
\newblock \bibinfo{journal}{\bibinfo{title}{Magnetization reversal driven by
  low dimensional chaos in a nanoscale ferromagnet}}.
\newblock {\emph{\JournalTitle{Nat. Commun.}}} \textbf{\bibinfo{volume}{10}},
  \bibinfo{pages}{543} (\bibinfo{year}{2019}).

\bibitem{williame19}
\bibinfo{author}{Williame, J.}, \bibinfo{author}{Accoily, A.~D.},
  \bibinfo{author}{Rontani, D.}, \bibinfo{author}{Sciamanna, M.} \&
  \bibinfo{author}{Kim, J.-V.}
\newblock \bibinfo{journal}{\bibinfo{title}{Chaotic dynamics in a macrospin
  spin-torque nano-oscillator with delayed feedback}}.
\newblock {\emph{\JournalTitle{Appl. Phys. Lett.}}}
  \textbf{\bibinfo{volume}{114}}, \bibinfo{pages}{232405}
  (\bibinfo{year}{2019}).

\bibitem{taniguchi19}
\bibinfo{author}{Taniguchi, T.} \emph{et~al.}
\newblock \bibinfo{journal}{\bibinfo{title}{Chaos in nanomagnet via feedback
  current}}.
\newblock {\emph{\JournalTitle{Phys. Rev. B}}} \textbf{\bibinfo{volume}{100}},
  \bibinfo{pages}{174425} (\bibinfo{year}{2019}).

\bibitem{taniguchi19JMMM}
\bibinfo{author}{Taniguchi, T.}
\newblock \bibinfo{journal}{\bibinfo{title}{Synchronized, periodic, and chaotic
  dynamics in spin torque oscillator with two free layers}}.
\newblock {\emph{\JournalTitle{J. Magn. Magn. Mater.}}}
  \textbf{\bibinfo{volume}{483}}, \bibinfo{pages}{281--292}
  (\bibinfo{year}{2019}).

\bibitem{williame20}
\bibinfo{author}{Williame, J.} \& \bibinfo{author}{Kim, J.-V.}
\newblock \bibinfo{journal}{\bibinfo{title}{Effects of delayed feedback on the
  power spectrum of spin-torque nano-oscillators}}.
\newblock {\emph{\JournalTitle{J. Phys. D: Appl. Phys.}}}
  \textbf{\bibinfo{volume}{53}}, \bibinfo{pages}{49501} (\bibinfo{year}{2020}).

\bibitem{kamimaki21}
\bibinfo{author}{Kamimaki, A.} \emph{et~al.}
\newblock \bibinfo{journal}{\bibinfo{title}{Chaos in spin-torque oscillator
  with feedback circuit}}.
\newblock {\emph{\JournalTitle{Phys. Rev. Research}}}
  \textbf{\bibinfo{volume}{3}}, \bibinfo{pages}{043216} (\bibinfo{year}{2021}).

\bibitem{yamaguchi23}
\bibinfo{author}{Yamaguchi, T.}, \bibinfo{author}{Tsunegi, S.},
  \bibinfo{author}{Nakajima, K.} \& \bibinfo{author}{Taniguchi, T.}
\newblock \bibinfo{journal}{\bibinfo{title}{Computational capability for
  physical reservoir computing using a spin-torque oscillator with two free
  layers}}.
\newblock {\emph{\JournalTitle{Phys. Rev. B}}} \textbf{\bibinfo{volume}{107}},
  \bibinfo{pages}{054406} (\bibinfo{year}{2023}).

\bibitem{tsunegi23}
\bibinfo{author}{Tsunegi, S.} \emph{et~al.}
\newblock \bibinfo{journal}{\bibinfo{title}{Information {Processing} {Capacity}
  of {Spintronic} {Oscillator}}}.
\newblock {\emph{\JournalTitle{Adv. Intell. Syst.}}}
  \textbf{\bibinfo{volume}{5}}, \bibinfo{pages}{2300175}
  (\bibinfo{year}{2023}).

\bibitem{dieny16}
\bibinfo{editor}{Dieny, B.}, \bibinfo{editor}{Goldfarb, R.~B.} \&
  \bibinfo{editor}{Lee, K.-J.} (eds.) \emph{\bibinfo{title}{Introduction to
  Magnetic Random-Access Memory}} (\bibinfo{publisher}{Wiley-IEEE Press,
  Hoboken}, \bibinfo{year}{2016}).

\bibitem{lau16}
\bibinfo{author}{Lau, Y.-C.}, \bibinfo{author}{Betto, D.},
  \bibinfo{author}{Rode, K.}, \bibinfo{author}{Coey, J. M.~D.} \&
  \bibinfo{author}{Stamenov, P.}
\newblock \bibinfo{journal}{\bibinfo{title}{Spin-orbit torque switching without
  an external field using interlayer exchange coupling}}.
\newblock {\emph{\JournalTitle{Nat. Nanotechnol.}}}
  \textbf{\bibinfo{volume}{11}}, \bibinfo{pages}{758} (\bibinfo{year}{2016}).

\bibitem{taniguchi22srep}
\bibinfo{author}{Taniguchi, T.}, \bibinfo{author}{Ogihara, A.},
  \bibinfo{author}{Utsumi, Y.} \& \bibinfo{author}{Tsunegi, S.}
\newblock \bibinfo{journal}{\bibinfo{title}{Spintronic reservoir computing
  without driving current or magnetic field}}.
\newblock {\emph{\JournalTitle{Sci. Rep.}}} \textbf{\bibinfo{volume}{12}},
  \bibinfo{pages}{10627} (\bibinfo{year}{2022}).

\bibitem{brown63}
\bibinfo{author}{Jr, W. F.~Brown.}
\newblock \bibinfo{journal}{\bibinfo{title}{Thermal {F}luctuations of a
  {S}ingle-{D}omain {P}article}}.
\newblock {\emph{\JournalTitle{Phys. Rev.}}} \textbf{\bibinfo{volume}{130}},
  \bibinfo{pages}{1677} (\bibinfo{year}{1963}).

\bibitem{taniguchi20AIP}
\bibinfo{author}{Taniguchi, T.}
\newblock \bibinfo{journal}{\bibinfo{title}{Synchronization and chaos in spin
  torque oscillator with two free layers}}.
\newblock {\emph{\JournalTitle{AIP Adv.}}} \textbf{\bibinfo{volume}{10}},
  \bibinfo{pages}{015112} (\bibinfo{year}{2020}).

\end{thebibliography}
\end{document}